\documentclass{ws-ijgmmp}
\usepackage{iftex}
\ifTUTeX
\usepackage{fontspec}
\else
\usepackage[T1]{fontenc}
\usepackage[utf8]{inputenc} 
\DeclareUnicodeCharacter{200B}{{\hskip 0pt}}
\fi
\usepackage[colorlinks=true]{hyperref}
\usepackage[numbers,sort&compress]{natbib}
\usepackage{float}
\usepackage[subrefformat=parens, caption=false]{subfig}
\begin{document}

\title{A class of Taub-NUT-scalar metrics via Ehlers transformations}

\author{{Ali Derekeh$^1$ \footnote{a.derekeh@ph.iut.ac.ir}}, {Behrouz Mirza$^1$ \footnote{b.mirza@iut.ac.ir}}, {Pouya Heidari$^2$ \footnote{pouya.heidari@ucalgary.ca}}, {Fatemeh Sadeghi$^1$ \footnote{fatemeh.sadeghi96@ph.iut.ac.ir}} and Reza Bahani$^1$ \footnote{bahanireza@gmail.com} \\}

\address{$^1$Department of Physics,\\ Isfahan University of Technology,\\ Isfahan 84156-83111, Iran}
\address{$^2$Department of Physics and Astronomy,\\ University of Calgary,\\ Calgary Alberta T2N 1N4 Canada}

\maketitle


\begin{abstract}
We derive a class of Taub-NUT metrics in the presence of a scalar field (TNS) by using Ernst equations and potential, as well as using Ehlers transformations on the exact solutions that was recently introduced in \cite{Mirza2023, Mirza2023-2}. Furthermore, we investigate the effective potential, geodesics, topological charge, quasinormal modes (QNMs) and the deflection angle of light in a gravitational lensing for the obtained class of TNS metrics. We also use conformal transformations to generate a new class of exact solutions of the Einstein-conformal-scalar theory by using the obtained TNS solutions as seed metrics. Finally we compare QNMs of the class of exact solutions.
\end{abstract}


\maketitle
\section{Introduction}\label{sec1}
The singularities of general relativity are divided into two categories, black holes and naked singularities, depending on whether they are hidden in an event horizon or not. Although cosmic censorship conjecture (CCC) \cite{Penrose1969-2, Hawking-1} suggest that wherever there is a singularity in curved space-time, this singularity is covered by a special hypersurface called the event horizon, however, this issue has not been proved so far. While the theoretical properties of black holes have been extensively studied, the naked singularity solutions in general relativity that arise from dynamical collapse studies are still relatively poorly understood. The study of naked singularities can be useful in many ways. For example, the $\gamma$-metric (Zipoy-Voorhees metric) \cite{Darmois1927, erez1959gravitational, Zipoy1966, Voorhees1970}, is a vacuum solution of Einstein's field equations and includes a naked singularity. It has been suggested in \cite{Destounis2023, Clavijo2023} that the $\gamma$-metric describes many features of sgr-$A^*$ at the center of our galaxy including the deviation from the perfect spherical state and its shadow. For more studies on the gamma metric, see Refs. \cite{herrera2000geodesics, richterek2002einstein, chakrabarty2018unattainable, abdikamalov2019black, toshmatov2019harmonic, benavides2019charged, allahyari2019quasinormal, allahyari2020quasinormal, chakrabarty2022effects, hajibarat2022gamma, li2022constraining, chakrabarty2023constraining}. Also, naked singularities form a region with strong gravity and provide excellent laboratories to study quantum gravity and may have interesting astrophysical effects at high energies in the universe \cite{harada2002physical}. For these reasons, we have strong motivation to derive new exact solutions of the Einsteins equations.\\

Another spacetime that has a naked singularity is the Fisher-Janis-Newman-Winicour (FJNW) metric \cite{Fisher1948, Janis1968, Wyman1981}. The FJNW metric is an asymptotically flat and static spacetime in the presence of a scalar field. So far, many studies have been done on the FJNW metric, for example see \cite{Virbhadra1998, Dey2008, Chowdhury2012, Turimov2018} and references therein. Recently, in \cite{Mirza2023}, a class of three-parameter metrics in the presence of a scalar field have been introduced, which are transformed into FJNW and $\gamma$-metrics at certain values ​​of the parameters. In \cite{Mirza2023-2}, rotating form of the class of three parameter metrics is introduced. In this paper, we are going to derive a class of Taub-NUT metrics in the presence of a scalar field. In the following, we briefly describe the history of the emergence of gravomagnetic monopole in Einstein's field equations. A spacetime with gravomagnetic monopole, known as Taub-NUT metric, was first introduced by Taub in 1951 as a homogeneous vacuum cosmological model \cite{taub1951empty}, and in 1963, Newman, Unti and Tamburino (NUT) derived this metric as a generalization of Schwarzschild spacetime \cite{newman1963empty, Kramer1983Exact}. Although the physical interpretation of this metric is very complex and has unique and strange features, it should be taken seriously as one of the solutions to Einstein's field equations. Until now, various physical interpretations have been suggested for Taub-NUT metric, for example, the explanations of Misner \cite{misner1963flatter, misner1967contribution} and Bonnor \cite{bonnor1969new} can be mentioned. Misner defined periodic time coordinates to interpret Taub-NUT metric and eliminate the string-like defect that is physically unfamiliar and confusing. Bonnor described the Taub-NUT metric as a rotating massless rod. The model presented by him has two problems, firstly, it is not really massless, and secondly, there is an infinite angular momentum in his description, while the angular momentum of the Taub-NUT metric is zero \cite{israel1977line, manko2005physical}. Manko and Ruiz \cite{manko2005physical} described the NUT solution as the exterior field of two semi-infinite sources with negative mass that rotate in opposite direction and are attached to the poles of a finite static rod with positive mass. So far, many studies have been done on NUT and Taub-NUT metrics, for example, see \cite{gonzalez2018new, hennigar2019thermodynamics, bordo2019first, bordo2019thermodynamics, bordo2019misner, zhang2021nut, arratia2021hairy, barrientos2022gravitational, durka2022first, cano2022quasinormal}.\\

To obtain a class of Taub-NUT metrics in the presence of a scalar field, we use three different methods. In the first one, we use the Ernst method to solve Einstein equations \cite{Ernst1968-1, Ernst1968-2}. Depending on the space-time symmetries, Ernst's potential can be defined in different forms. For example, the authors in \cite{Mirza2023-2} have used a special type of Ernst potential to obtain a class of rotating metrics in the presence of a scalar field. In \cite{reina1976axisymmetric, kinnersley1977symmetries1, kinnersley1977symmetries, kinnersley1978symmetries1, kinnersley1978symmetries, hoenselaers1979symmetries1, hoenselaers1979symmetries, cosgrove1980relationships, wu2005two}, some methods to derive Ernst potentials for different space-time symmetries are described in detail.\\ 

Another interesting way to obtain TNS metrics is to use the Ehlers transformations \cite{Ernst1968-3, Ernst1968-4, Ernst1976}. The Ehlers transformations are obtained according to the symmetry of Ernest equations. By using Ehlers transformations, it is possible to add the NUT parameter \cite{Misner1967, Chng2006} to axially symmetric metrics. Also, it is proved that the Ehlers transformations can be used for the Einstein-Hilbert action coupled with the scalar fields \cite{astorino2013embedding, astorino2020enhanced}. We consider this fact and derive a new class of Taub-NUT-scalar metrics by using Ehlers transformation. We show that the results obtained by using either Ernst method or Ehlers transformations are the same. To see some examples of the applications of Ehlers transformations , see Refs. \cite{barrientos2023ehlers, barrientos2023pleban, Cisterna2023}.\\ 

After obtaining a class of TNS metrics, following the Duran's theory \cite{duan1979TSU, duan1984structure, Duan2000}, a topological current and charge for light rings is obtained for an arbitrary member of the new class of Taub-NUT metrics and it is proved that there is at least one light ring whose topological number is -1. By using conformal transformations \cite{Bekenstein1947} and the new class of Taub-NUT metrics in the presence of a scalar field, we obtain a new class of exact solutions of the Einstein-conformal-scalar theory. The conformal transformations used here include a redefinition of the massless scalar field and a conformal rescaling of the metric. Additionally, we investigated the curvature invariants, singularities, quasinormal modes (QNMs) and gravitational lensing  for the TNS class of metrics.\\

This article is organized as follows. In Sec. \ref{sec:three-parameter}, we go through a quick preliminary review of the new class of metrics in the presence of a massless scalar field. Then, in Sec. \ref{sec:TNS}, we obtain the class of TNS metrics by using three different methods i.e. by introducing  the Ernst potential and two different algorithms for Ehlers transformations. Then, we investigate various properties of the TNS metric, such as geodesics, effective potential, and topological charge. In Sec. \ref{sec:Cthree}, we consider the Bekenstein approach to find the proper conformal transformation to derive a new class of exact solutions of the Einstein-conformal-scalar theory that are counterpart of the class of three parameter static metrics. Thereafter, in Sec. \ref{sec:CTNS}, we propose the conformal transformation of the class of TNS metrics, and obtain another new class of Taub-NUT metrics related to the Einstein-conformal-scalar theory. We also investigate the geodesics and the effective potential. Finally, in Sec. \ref{sec:QNM}, we derive the QNMs of TNS and conformal TNS (CTNS) metrics using the light ring method. In Sec. \ref{sec:Lensing}, the deflection angle of light due to gravitational lensing is investigated for TNS metrics. Sec. \ref{sec:con} is devoted to the conclusions.
\section{A CLASS OF METRICS IN THE PRESENCE OF A MASSLESS SCALAR FIELD}\label{sec:three-parameter}
Recently, a class of exact solutions of Einstein's equations and their rotating forms were introduced in \cite{Mirza2023, Mirza2023-2}. The class of static solutions depends on three parameters, i.e., mass and two other free parameters and one may obtain the well-known $\gamma$-metric (Zipoy-Voorhees metric) and the FJNW metric at some values of the parameters. In the following, we will consider the Einstein-Hilbert action ($ 8\,\pi\,G=c=1 $) in the presence of a massless scalar field
\begin{equation}\label{eq1.1}
S = \int d^4 x \sqrt{-g} \left(\mathcal{R} - \partial_{\alpha} \psi(r) \partial^{\alpha} \psi(r)\right)~,
\end{equation}
where $\mathcal{R}$, $g$, and $\psi$ are the Ricci scalar, determinant of metric, and scalar field, respectively. By varying the above action with respect to the metric tensor $g_{\mu \nu}$ and the massless scalar field $\psi$, respectively, the equations of motion will be obtained as follows
\begin{gather}
\mathcal{R_{\alpha \beta}} = \partial_{\alpha} \psi(r) \partial_{\beta} \psi(r)~,\label{eq1.2} \\ 
\nabla_{\alpha} \nabla^{\alpha} \psi(r)=0~. \label{eq1.3}
\end{gather}
Now, we consider the following exact solutions of the Einstein's equations that recently was obtained in \cite{Mirza2023}
\begin{equation}\label{eq1.4}
ds^2 = -f(r)^\gamma dt^2 + f(r)^\mu k(r, \theta)^\nu (\frac{dr^2}{f(r)} + r^2 d\theta^2) + f(r)^{1-\gamma} r^2 \sin^2 (\theta) d\phi^2~,
\end{equation}
where the functions $f(r)$ and $k(r,\theta)$ are given as below 
\begin{gather}
f(r)=1- \frac{2 m}{r}~,\label{eq1.5} \\ 
k(r, \theta) = 1- \frac{2 m}{r} + \frac{m^2}{r^2} \sin^2 (\theta)~. \label{eq1.6}
\end{gather}
The metric in \eqref{eq1.4} depends on three independent parameters. The parameters are mass $m$, and three other parameters $\gamma$, $\mu$, and $\nu$ that only two of them are independent as we have a constraint, i.e. $\mu+\nu=1-\gamma$. The expression for the scalar field $\psi$ can be written as
\begin{equation}\label{eq1.7}
\psi(r) = \sqrt{\frac{1- \gamma^2 - \nu}{2}} \ln (1- \frac{2 m}{r})~.
\end{equation}
Assuming that the scalar field is a real function, the following conditions on the parameters $\mu$ and $\nu$ can be obtained
\begin{gather}
\nonumber \mu \geq \gamma^2 - \gamma ~,\\ 
\nu\le 1- \gamma^2~. \label{eq1.8}
\end{gather}
Overall, this three parameter solution is important. Choosing the parameters as $\mu = \gamma^2 - \gamma$ and $\nu = 1- \gamma^2$, the metric represents the $\gamma$-metric, and by choosing $\nu=0$ will give us the FJNW metric. We will define the conformal transformation of metric \eqref{eq1.4} and hence derive a new exact solution for the Einstein-conformal-scalar theory in section \ref{sec:Cthree} by using the scalar field \eqref{eq1.7}. Various futures of the class of metrics in the presence of a scalar field have been investigated in \cite{Mirza2023, Mirza2023-2}.
\section{TAUB-NUT-SCALAR METRICS}
In this section, we intend to introduce a class of Taub-NUT metrics in the presence of a scalar field. To derive the class of metrics, we use three methods, the first method is to solve the Einstein field equations by directly introducing the Ernst potential, and then, we use Ehlers transformations by two approaches.
\subsection{TAUB-NUT-SCALAR METRICS VIA ERNST POTENTIAL}
For the calculations related to this section, we first consider the Lewis-Weyl-Papapetrou (LWP) metrics in the $ (t,\,\rho,\,z,\,\phi) $ coordinates
\begin{equation}\label{500_label}
ds^2 = - f \, (d t - \omega \, d \phi)^2 + \frac{\sigma^2}{f} \, \big[ e^{2 \, \lambda} \, (d \rho^2 + d z^2) + \rho^2 \, d \phi^2 \big],
\end{equation}
where sigma is a positive parameter and $f$, $\omega$ and $\lambda$ are functions of $\rho$ ​​and $z$. Next, our goal is to find the form of these functions. First, according to Eq. \eqref{eq1.2} and metric \eqref{500_label}, we have
\begin{equation}\label{502_label}
R_{tt}=0,\;\Rightarrow\;
f \, (\partial_\rho^2 \, f + \partial_z^2 \, f + \frac{\partial_\rho \, f}{\rho}) - (\partial_\rho \, f)^2 - (\partial_z \, f)^2+ \frac{f^4}{\rho^2 \, \sigma^2} \; \big[ (\partial_\rho \, \omega)^2 + (\partial_z \, \omega)^2 \big]  = 0,
\end{equation}
\begin{equation}\label{504_label}
R_{tt}+R_{t\phi}=0,\quad\Rightarrow\quad\partial_\rho \, (\frac{f^2 \, \partial_\rho \, \omega}{\rho}) + \partial_z  \, (\frac{f^2 \, \partial_z \, \omega}{\rho}) = 0.
\end{equation}
Now we define a function $ \chi $ as follows
\begin{equation}\label{505_label}
\partial_\rho \, \chi = - \frac{f^2 \, \partial_z \, \omega}{\rho \, \sigma}, \qquad \partial_z \, \chi =  \frac{f^2 \, \partial_\rho \, \omega}{\rho \, \sigma}.
\end{equation}
Using the definition \eqref{505_label}, we rewrite Eq. \eqref{502_label} in the following form
\begin{equation}\label{510_label}
f \, \nabla^2 \, f - (\partial_\rho \, f)^2 - (\partial_z \, f)^2 + (\partial_\rho \, \chi)^2 +  (\partial_z \, \chi)^2 = 0,
\end{equation}
where $ \nabla^2 \, f $ is defined as follows
\begin{equation}\label{511_label}
\nabla^2 \, f  = \partial_\rho^2 \, f + \partial_z^2 \, f + \frac{\partial_\rho \, f}{\rho}.
\end{equation}
Using Eqs. \eqref{505_label} and \eqref{511_label}, we write the following relation for $ \chi $
\begin{equation}\label{512_label}
f \, \nabla^2 \, \chi = 2 \, (\partial_\rho \, f \, \partial_\rho \, \chi + \partial_z \, f \, \partial_z \, \chi).
\end{equation}
Using Eq. \eqref{eq1.2} and metric \eqref{500_label}, we calculate $ R_{\rho\rho} $ and $ R_{zz} $ and after simplifying and using Eq. \eqref{505_label} we get the following equation
\begin{equation}\label{507_label}
\begin{aligned}
&R_{\rho\rho}-R_{zz}=(\partial_\rho \, \psi)^2 - (\partial_z \, \psi)^2,\\
&\Rightarrow\,\partial_\rho \, \lambda = \frac{\rho}{4} \, \big\{ \frac{1}{f^2} \, \big[ (\partial_\rho \, f )^2 - (\partial_z \, f)^2 + (\partial_\rho \, \chi)^2 - (\partial_z \, \chi)^2 \big]+ 2 \, [(\partial_\rho \, \psi)^2 - (\partial_z \, \psi)^2] \big\}.
\end{aligned}
\end{equation}
Using Eq. \eqref{eq1.2}, metric \eqref{500_label}, Eq. \eqref{505_label} and $ R_{\rho z} $, we have
\begin{equation}\label{506_label}
R_{\rho z}=\partial_\rho \, \psi \, \partial_z \, \psi\quad\Rightarrow\quad\partial_z \, \lambda = \frac{\rho}{2} \, \big[ \frac{1}{f^2} \, (\partial_\rho \, f \, \partial_z \, f + \partial_\rho \, \chi \, \partial_z \, \chi) + 2 \, \partial_\rho \, \psi \, \partial_z \, \psi \big].
\end{equation}
Now using the following coordinate transformations
\begin{equation}\label{500a_label}
\begin{aligned}
& t = t, \quad \phi = \phi, \quad \rho = (x^2 - 1)^{\frac{1}{2}} \; (1 - y^2)^{\frac{1}{2}}, \quad z = x \, y,\\
& x = \frac{1}{2} \, (R_+ + R_-), \quad y = \frac{1}{2} \, (R_+ - R_-),\qquad R_\pm = \sqrt{\rho^2 + (z \pm 1)^2}.
\end{aligned}
\end{equation}
We can rewrite Eqs. \eqref{505_label}, \eqref{507_label} and \eqref{506_label} in the $ (t,\,x,\,y,\,\phi) $ coordinates as follows
\begin{equation}\label{24_label}
\partial_x \, \omega = \sigma \, (1 - y^2) \, f^{-2} \; \partial_y \chi, \quad \partial_y \, \omega = \sigma \, (1 - x^2) \, f^{-2} \; \partial_x \chi,
\end{equation}
\begin{equation}\label{25_label}
\begin{aligned}
\partial_x \, \lambda & = \frac{(1 - y^2) \, f^{- 2}}{4 \, (x^2 - y^2)} \, \big\{ x \, (x^2 - 1) \, \big[ (\partial_x \, f)^2 + (\partial_x \, \chi)^2 \big]+ x \, (y^2 - 1) \, \big[ (\partial_y \, f)^2 + (\partial_y \, \chi)^2 \big]\\
& - 2 \, y \, (x^2 - 1) \, (\partial_x \, f \, \partial_y \, f + \partial_x \, \chi \, \partial_y \, \chi)  \big\}+\frac{(1 - y^2)}{2\,(x^2 - y^2)} \, x \, (x^2 - 1) \, (\partial_x \, \psi)^2,
\end{aligned}
\end{equation}
\begin{equation}\label{26_label}
\begin{aligned}
\partial_y \, \lambda = & \frac{(x^2 - 1) \, f^{- 2}}{4 \, (x^2 - y^2)} \, \big\{ y \, (x^2 - 1) \, \big[ (\partial_x \, f)^2 + (\partial_x \, \chi)^2 \big]- y \, (1 - y^2) \, \big[ (\partial_y \, f)^2 + (\partial_y \, \chi)^2 \big]\\
& + 2 \, x \, (1 - y^2) \, (\partial_x \, f \, \partial_y \, f + \partial_x \, \chi \, \partial_y \, \chi)  \big\}+\frac{(x^2 - 1)^2 \, y}{2\,(x^2 - y^2)} \, (\partial_x \, \psi)^2.
\end{aligned}
\end{equation}
Also, metric \eqref{500_label} in the $ (t,\,x,\,y,\,\phi) $ coordinates is written as follows
\begin{equation}\label{8_label}
d s^2 = - f \, (dt - \omega \, d \phi)^2 + \frac{\sigma^2}{f} \, \big[ e^{2 \, \lambda} \, (x^2 - y^2)( \frac{d x^2}{x^2 - 1} + \frac{d y^2}{1 - y^2} ) + (x^2 - 1) \, (1 - y^2) \, d \phi^2 \big].
\end{equation}
By defining the complex Ernst potential $ \varepsilon $ as follows
\begin{equation}\label{18_label}
\varepsilon = f + i \, \chi.
\end{equation}
We can represent Eqs. \eqref{510_label} and \eqref{512_label} with the following Ernst equation
\begin{equation}\label{17_label}
(\varepsilon + \varepsilon^*) \, \Delta \varepsilon = 2 \, (\nabla \varepsilon)^2,
\end{equation}
where $ \varepsilon^* $ is the complex conjugate of $ \varepsilon $ and $\nabla$ and $\Delta$ are defined as follows
\begin{equation}\label{19_label}
\begin{aligned}
\nabla \equiv & \,  \sigma^{- 1} \, (x^2 - y^2)^{- \frac{1}{2}} \; {\hat i \, \big[ (x^2 - 1)^{\frac{1}{2}} \, \partial_x \big]  + \hat j \, \big[ (1 - y^2)^{\frac{1}{2}} \, \partial_x \big]},\\
\Delta \equiv & \,  \sigma^{- 2} \, (x^2 - y^2)^{- 1} \; {\partial_x \, \big[ (x^2 - 1) \, \partial_x \big] + \partial_x \, \big[ (x^2 - 1) \, \partial_x \big]}.
\end{aligned}
\end{equation}
The following Ernst potential function can explain our stationary solution. The following Ernst potential can also be used to obtain the Taub-NUT metric. We can use the same Ernst potential here since the scalar field does not affect the Ernst equation and therefore the Ernst potential can be written as
\begin{equation}\label{20_label}
\varepsilon = \frac{\sigma\,x-m-i\,n}{\sigma\,x+m+i\,n},
\end{equation}
where $n$ is the NUT parameter and $\sigma$ is defined as follows
\begin{equation}\label{22_label}
\sigma=\sqrt{m^2+n^2}.
\end{equation}
Using Eqs. \eqref{24_label}, \eqref{18_label}, \eqref{20_label} and \eqref{22_label}, we obtain $ f $, $ \chi $ and $ \omega $ as follows
\begin{equation}\label{220_label}
f=\frac{\sigma\,(x^2-1)}{2\,m\,x+\sigma\,(x^2+1)},\qquad\chi=-\frac{2\,n\,x}{2\,m\,x+\sigma\,(x^2+1)},\qquad \omega=-2\,n\,y.
\end{equation}
It is clear that derivative of Eq. \eqref{25_label} with respect to $ y $, is equal to derivative of Eq. \eqref{26_label} with respect to $ x $. By using $ \partial_y\,\partial_x\,\lambda=\partial_x\,\partial_y\,\lambda $ we obtain the following differential equation
\begin{equation}\label{27_label}
\psi^{\prime \prime} + \frac{2 \, x}{x^2 - 1} \, \psi^{\prime} = 0, \quad \Rightarrow \quad \psi = c_1 \, \ln \big( \frac{x - 1}{x + 1} \big) + c_2.
\end{equation}
Using Eq. \eqref{eq1.3} and metric \eqref{8_label}, we obtain the coefficients of $ c_1 $ and $ c_2 $ as follows
\begin{equation}\label{28_label}
c_1 = \sqrt{\frac{- \nu}{2}}, \qquad c_2 = 0.
\end{equation}
Now by putting Eqs. \eqref{24_label}, \eqref{220_label}, \eqref{27_label} and \eqref{28_label} into Eqs. \eqref{25_label} and \eqref{26_label} and integrating the obtained relations, we have
\begin{equation}\label{11_label}
e^{2 \lambda} =\big( \frac{x^2 - 1}{x^2 - y^2} \big)^{1 - \nu}.
\end{equation}
Finally, using the following coordinate transformation
\begin{equation}\label{14_label}
x = \frac{r - m}{\sqrt{m^2 + n^2}}, \qquad y = cos \theta.
\end{equation}
and using Eqs. \eqref{220_label} and \eqref{11_label}, we derive the final form of the Taub-NUT-scalar metric as follows
\begin{equation}\label{h18_label}
\begin{aligned}
&ds^2 = -\left(1 - \frac{2 \left(n^2 + m r\right)}{r^2 + n^2}\right) dt^2 + \frac{\left(r^2 + n^2\right) \left(1+ \frac{\left(m^2 + n^2\right) \sin^2 (\theta)}{r^2 - 2 m r - n^2} \right)^{\nu}}{r^2 - 2 m r - n^2} dr^2\\
& + \left(r^2 + n^2\right) \left(1 + \frac{\left(m^2 + n^2\right) \sin^2 (\theta)}{r^2 - 2 m r - n^2}\right)^{\nu} d\theta^2 \\
&+ \left(- \frac{ 4 n^2 \left(r^2 - 2 m r - n^2\right) \cos^2 (\theta)}{r^2 + n^2} + \left(r^2 + n^2 \right)\sin^2(\theta) \right) d\phi^2\\
& - \frac{4 n \left(r^2 - 2 m r - n^2\right) \cos(\theta)}{r^2 + n^2} dt d\phi~. 
\end{aligned}
\end{equation}
Also, using coordinate transformation \eqref{14_label}, the scalar field \eqref{27_label} is written as follows
\begin{equation}\label{h190_label}
\psi (r) =\sqrt{\frac{- \nu}{2}} \ln \left(\frac{r - m - \sqrt{m^2 + n^2}}{r - m + \sqrt{m^2 + n^2}}\right)~.
\end{equation}
It is interesting that we can derive the class of TNS metrics by using two different but equivalent form of  Ehlers transformations. 
\subsection{TAUB-NUT-SCALAR METRIC VIA EHLERS TRANSFORMATION}\label{sec:TNS}
In the following we use the class of three parameter metrics \eqref{eq1.4} and apply the Ehlers transformation to drive a class of Taub-NUT-Scalar (TNS) metrics. There are two different but equivalent methods to do this. Each of these methods is explained below. In both methods, we assume that $\gamma=1$ and $\mu=- \nu$ in \eqref{eq1.4}. In the calculations related to this section, we should note that although the Ehlers transformations are written in vacuum, they can be used for the Einstein-Hilbert action coupled with the scalar field. In this case, the solutions of Ernst's equations remain unchanged and an equation related to the scalar field is as was proved in \cite{astorino2013embedding, astorino2020enhanced}.
\subsubsection{The first method of Ehlers transformations}\label{sec3.2.1}
In order to be able to use Ehlers transformations, we must first write the metric \eqref{eq1.4} in the form of Lewis-Weyl-Papapetrou (LWP) metric as follows
\begin{equation}\label{h1_label}
ds^2=-f\,(dt-\omega\,d\phi)^2+f^{-1}\,\big[e^{2\,\lambda}\,(d\rho^2+dz^2)+\rho^2\,d\phi^2\big],
\end{equation}
In the following, we first use of the following transformation coordinates in metric \eqref{h1_label}
\begin{equation}\label{h2_label}
\rho=\sqrt{r\,(r-2\,m)}\,\sin\theta, \qquad z=(r-m)\,\cos\theta.
\end{equation}
And then by comparing with metric \eqref{eq1.4} (in $\gamma=1$), we obtain $ f $, $ \omega $ and $ \lambda $ functions as follows
\begin{equation}\label{h3_label}
\begin{aligned}
&f(r) = 1 - \frac{2\,m}{r},\\
&\omega=0,\\
&e^{2\,\lambda}=\big[1+\frac{m^2\,\sin^2\theta}{r\,(r-2\,m)}\big]^\nu\;\frac{r\,(r-2\,m)}{r^2-2\,m\,r+m^2\,\sin^2\theta}.
\end{aligned}
\end{equation}
In \cite{Ernst1968-1, Ernst1968-2}, Ernst introduced the following complex potential
\begin{equation}\label{h4_label}
\varepsilon=f+i\,\chi,
\end{equation}
where $ \chi $ is defined as follows
\begin{equation}\label{h6_label}
\hat{\phi}\times\nabla\,\chi:=-\rho^{-1}\,f^2\,\nabla\,\omega,
\end{equation}
where $ \hat{\phi} $ represents the unit vector in the $ \phi $ direction. Now, we introduce the Ehlers transformations
\begin{equation}\label{h8_label}
\varepsilon^\prime = \frac{\varepsilon}{1+i\,c\,\varepsilon},
\end{equation}
where $ c $ is a real parameter. According to metric \eqref{eq1.4} in the $ \gamma=1 $ and using the Eqs. \eqref{h3_label}, \eqref{h4_label} and \eqref{h6_label}, one has
\begin{equation}\label{h9_label}
\varepsilon=1 - \frac{2\,m}{r}, \qquad \chi=0.
\end{equation}
By placing Eq. \eqref{h9_label} in Eqs. \eqref{h8_label}, we have
\begin{equation}\label{h10_label}
\varepsilon^\prime=\frac{r\,(r-2\,m)}{r^2+c^2\,(r-2\,m)^2}-i\,\frac{c\,(r-2\,m)^2}{r^2+c^2\,(r-2\,m)^2}.
\end{equation}
Using Eqs. \eqref{h4_label} and \eqref{h10_label}, we obtain
\begin{equation}\label{h11_label}
f^\prime=\frac{r\,(r-2\,m)}{r^2+c^2\,(r-2\,m)^2},\qquad\chi^\prime=-\frac{c\,(r-2\,m)^2}{r^2+c^2\,(r-2\,m)^2}.
\end{equation}
Now we need to get $\omega^\prime$ using Eq. \eqref{h6_label}. In Eq. \eqref{h6_label}, the gradients are written in the cylindrical coordinate system, and we use the following equation to convert them to the spherical coordinate system
\begin{equation}\label{h14_label}
\nabla\,F(r,\,\theta)=\frac{1}{\sqrt{(r-m)^2-m^2\,\cos^2\theta}}\;\big[\frac{\partial\,F(r,\,\theta)}{\partial\,r}\,\sqrt{r\,(r-2\,m)}\,\hat{r}+\frac{\partial\,F(r,\,\theta)}{\partial\,\theta}\,\hat{\theta}\,\big].
\end{equation}
According to Eqs. \eqref{h2_label}, \eqref{h6_label}, \eqref{h10_label}, \eqref{h11_label} and \eqref{h14_label}, we obtain $\omega^\prime$ as follows
\begin{equation}\label{h15_label}
\omega^\prime=4\,c\,m\,\cos \theta.
\end{equation}
By placing Eqs. \eqref{h2_label}, \eqref{h3_label}, \eqref{h10_label} and \eqref{h15_label} in metric \eqref{h1_label}, we have
\begin{equation}\label{h16_label}
\begin{aligned}
ds^2&=-\frac{r\,(r-2\,m)}{r^2+c^2\,(r-2\,m)^2}\,(dt-4\,c\,m\,\cos \theta\,d\phi)^2+\frac{r^2+c^2\,(r-2\,m)^2}{r\,(r-2\,m)}\\
&\times\big[(1+\frac{m^2\,\sin^2\theta}{r\,(r-2\,m)})^\nu\,\big(dr^2+r\,(r-2\,m)\,d\theta^2\big)\big]+\big(r^2+c^2\,(r-2\,m)^2\big)\,\sin^2\theta\,d\phi^2.
\end{aligned}
\end{equation}
Now we use the following coordinate transformations
\begin{equation}\label{h17_label}
\begin{aligned}
&\tilde{r}=r\,\sqrt{1+c^2}-\frac{2\,m\,c^2}{\sqrt{1+c^2}},\qquad \tilde{t}=\frac{t}{\sqrt{1+c^2}}\\ &m=-\frac{n}{2\,c}\,\sqrt{1+c^2},\qquad c=\frac{\tilde{m}-\sqrt{\tilde{m}^2+n^2}}{n},
\end{aligned}
\end{equation}
Finally, using transformations \eqref{h17_label} in metric \eqref{h16_label}, and remove the tilde ($\tilde{t}\rightarrow t$, $\tilde{r}\rightarrow r$ and $\tilde{m}\rightarrow m$) we derive the final form of the Taub-NUT-Scalar metric as follows
\begin{equation}\label{h180_label}
\begin{aligned}
&ds^2 = -\left(1 - \frac{2 \left(n^2 + m r\right)}{r^2 + n^2}\right) dt^2 + \frac{\left(r^2 + n^2\right) \left(1+ \frac{\left(m^2 + n^2\right) \sin^2 (\theta)}{r^2 - 2 m r - n^2} \right)^{\nu}}{r^2 - 2 m r - n^2} dr^2\\
& + \left(r^2 + n^2\right) \left(1 + \frac{\left(m^2 + n^2\right) \sin^2 (\theta)}{r^2 - 2 m r - n^2}\right)^{\nu} d\theta^2 \\
&+ \left(- \frac{ 4 n^2 \left(r^2 - 2 m r - n^2\right) \cos^2 (\theta)}{r^2 + n^2} + \left(r^2 + n^2 \right)\sin^2(\theta) \right) d\phi^2\\
& - \frac{4 n \left(r^2 - 2 m r - n^2\right) \cos(\theta)}{r^2 + n^2} dt d\phi~. 
\end{aligned}
\end{equation}
Using the metric \eqref{h180_label} and the relation $ \Box\,\psi(r)=0 $, we obtain the scalar field $ \psi(r) $ as follows
\begin{equation}\label{h19_label}
\psi (r) =\sqrt{\frac{- \nu}{2}} \ln \left(\frac{r - m - \sqrt{m^2 + n^2}}{r - m + \sqrt{m^2 + n^2}}\right)~.
\end{equation}
The folllowing coordinate transformations can be used to derive TNS metric in \eqref{h180_label} from the LWP in \eqref{h1_label}
\begin{equation}\label{h20_label}
\begin{aligned}
&f=\frac{(R_++R_-)^2-4\,(m^2+n^2)}{(R_++R_-+2\,m)^2+4\,n^2},\qquad \omega=-\frac{n}{\sqrt{m^2+n^2}}\,(R_+-R_-),\\ &e^{2\,\gamma}=\big(\frac{(R_++R_-)^2-4\,(m^2+n^2)}{4\,R_+\,R_-}\big)^{1-\nu},\qquad R_{\pm}^2=r-m\pm\sqrt{m^2+n^2}\,\cos\theta,
\end{aligned}
\end{equation}
where $ \rho $ and $ z $ in \eqref{h1_label} are defined as below
\begin{equation}\label{h21_label}
\rho=\sqrt{r^2-2\,m\,r-n^2}\,\sin\theta,\qquad z=(r-m)\,\cos\theta.
\end{equation}
In the following, we use another type of Ehlers transformations to obtain the class of TNS metrics.
\subsubsection{The second method of Ehlers transformations}\label{sec3.2.2}
\begin{figure}[H]
\centering
\includegraphics[width=0.7\linewidth]{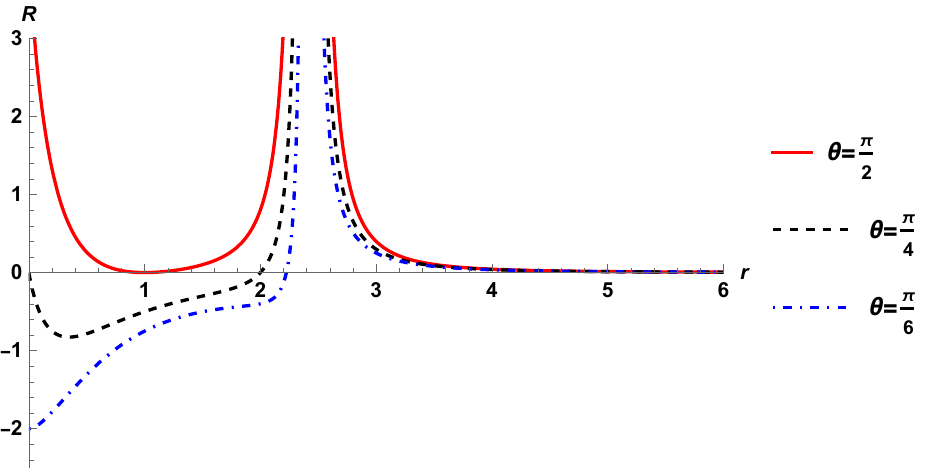}
\caption{The Ricci scalar for a TNS metric with parameters $m=1$, $n=1$, $\nu=-1$, diverges at $r=1+\sqrt{2}$ for different values of $\theta$.}
\label{fig:nutricci}
\end{figure}

It is interesting that by using another type of Ehlers transformations \cite{Ehlers1962, Momeni2005, Alawadhi2020} we can derive the TNS class of metrics. We write the metric in Eq. \eqref{eq1.4} as follows
\begin{equation}\label{eq1.9}
ds^2 = -f(r) dt^2 + f(r)^\mu k(r, \theta)^{\nu} (\frac{dr^2}{f(r)} + r^2 d\theta^2) + r^2 \sin^2 (\theta) d\phi^2~.
\end{equation}
The metric in \eqref{eq1.9} has at least one time-like killing vector representing the isometry. We define the 4-dimensional line element as follows ($x^\mu=\{x^0,\, x^i\}$)
\begin{equation}\label{eq1.10}
ds^2 = -e^{2 U} (dx^0 + A_{i} dx^{i})^2 + dl^2~,
\end{equation}
where
\begin{gather}
A_{i} = -\frac{g_{0 i}}{g_{0 0}}, \hspace{5pt} e^{2 U} = - g_{0 0}, \hspace{5pt} dl^2 = \gamma_{i j} dx^{i} dx^{j}~,\label{eq1.11} \\
\gamma_{i j} = - g_{ij} + \frac{g_{0 i} g_{0 j}}{g_{0 0}}~. \label{eq1.12}
\end{gather}
The Ehlers transformation states that if the space-time can be written in the form
\begin{equation}\label{eq1.13}
g_{\mu\nu}\,dx^\mu\, dx^\nu = -e^{2 U}(dx^{0})^2 + e^{-2 U} d\tilde{l}^2,
\end{equation}
where $d\tilde{l}^2 = e^{2 U} dl^2$, we can find a new stationary solution that satisfies Einstein's field equations as follows
\begin{equation}\label{eq1.14}
\bar{g}_{\mu \nu}\, dx^{\mu}\, dx^{\nu} = -\left(\alpha \cosh(2 U)\right)^{-1} \left(dx^{0} + A_{i} dx^{i}\right)^2 - \alpha \cosh(2 U) d\tilde{l}^2~,
\end{equation}
where $\alpha$ has a constant and positive value and the functions $U=U(x^{i})$ and $A_{i}(x^{j})$ satisfy the Ehlers equation
\begin{equation}\label{eq1.15}
-\alpha \sqrt{\tilde{\gamma}} \epsilon_{i j k} U^{, k} = A_{\left[i, j\right]}~.
\end{equation}
where $\bar{\gamma}$ is determinant of the spatial part of the conformal metric and $A_{\left[i, j\right]} = \frac{\partial A_{i}}{\partial x^{j}} - \frac{\partial A_{j}}{\partial x^{i}}$. According to the Ehlers approach, for the metric \eqref{eq1.4}, the function $ U $ can be defined as
\begin{equation}\label{eq1.16}
U_{c} = \frac{1}{2} \ln \left(1- \frac{2 m}{r}\right) + \frac{1}{2} \ln(c)~,
\end{equation}
where $c$ is a constant. As a result, for the three-parameter metric with $\gamma=1$, we have
\begin{equation}\label{eq1.17}
d\tilde{l}^2 = \left(1+ \frac{m^2 \sin^2 (\theta)}{r^2 - 2 m r}\right)^{\nu} dr^2 + r^2 \left(1- \frac{2 m}{r}\right) \left[\left(1+ \frac{m^2 \sin^2 (\theta)}{r^2 - 2 m r}\right)^{\nu} d\theta^2 + \sin^2 (\theta) d\phi^2\right]~,
\end{equation}
and the determinant of the metric is
\begin{equation}\label{eq1.18}
\tilde{\gamma} = r^4 \left(1 - \frac{2 m}{r}\right)^{2} \left(1+ \frac{m^2 \sin^2 (\theta)}{r^2 - 2 m r}\right)^{2 \nu} \sin^2 (\theta)~.
\end{equation}
If we choose the field $A_{i}$ in such a way that it is only a function of $\theta$, by solving equation \eqref{eq1.15} and using metric \eqref{eq1.17}, we get
\begin{equation}\label{eq1.19}
A_{\phi}(\theta) = -2 \alpha~m \cos(\theta)~.
\end{equation}
By substituting the field \eqref{eq1.19} in \eqref{eq1.14}, it is just a matter of calculation to find
\begin{equation}\label{eq1.20}
\begin{aligned}
& ds^2 = - \frac{2 c \left(r^2 - 2 m r\right)}{\alpha \left[r^2 + c^2 \left(r - 2 m\right)^2\right]} dt^2  - \frac{ 8 c m \left(r^2 - 2 m r\right) \cos(\theta)}{r^2 + c^2 \left(r - 2 m\right)^2 } dt d\phi\\
& + \frac{\alpha \left(1+ c^2 \left(1- \frac{2 m}{r}\right)^{2} \right) \left(1+ \frac{m^2 \sin^2 (\theta)}{r^2 - 2 m r}\right)^{\nu}}{2 c \left(1- \frac{2 m}{r}\right)} dr^2 \\
&+ \frac{\alpha \left[r^2 + c^2 \left(r- 2 m\right)^2\right] \left(1+ \frac{m^2 \sin^2 (\theta)}{r^2 - 2 m r}\right)^{\nu}}{2 c} d\theta^2\\
& + \alpha \left(- \frac{8 c m^2 r \left(r - 2 m\right) \cos^2 (\theta)}{r^2 + c^2 \left(r - 2 m\right)^2} + \frac{\left[r^2 + c^2 \left(r - 2 m\right)^2\right]\sin^2 (\theta)}{2 c}\right) d\phi^2~.
\end{aligned}
\end{equation}
Now, if we make the following substitutions 
\begin{equation}\label{eq.sub}
\begin{aligned}
& r = R-M+\sqrt{M^2+n^2}, \hspace{10pt} \alpha= \frac{n}{\sqrt{M^2 +  n^2}}~,\\
& m = \sqrt{M^2+  n^2}, \hspace{10pt} c=\frac{\sqrt{M^2  +n^2}  -M}{n}~,
\end{aligned}
\end{equation}
in (\ref{eq1.20}) where $n$ is the NUT charge, and then using the following replacements $M \rightarrow m$ and $R \rightarrow r$, we will find the class of TNS metrics as below
\begin{equation}\label{eq1.21}
\begin{aligned}
&ds^2 = -\left(1 - \frac{2 \left(n^2 + m r\right)}{r^2 + n^2}\right) dt^2 + \frac{\left(r^2 + n^2\right) \left(1+ \frac{\left(m^2 + n^2\right) \sin^2 (\theta)}{r^2 - 2 m r - n^2} \right)^{\nu}}{r^2 - 2 m r - n^2} dr^2\\
& + \left(r^2 + n^2\right) \left(1 + \frac{\left(m^2 + n^2\right) \sin^2 (\theta)}{r^2 - 2 m r - n^2}\right)^{\nu} d\theta^2 \\
&+ \left(- \frac{ 4 n^2 \left(r^2 - 2 m r - n^2\right) \cos^2 (\theta)}{r^2 + n^2} + \left(r^2 + n^2 \right)\sin^2(\theta) \right) d\phi^2\\
& - \frac{4 n \left(r^2 - 2 m r - n^2\right) \cos(\theta)}{r^2 + n^2} dt d\phi~. 
\end{aligned}
\end{equation}
Now, considering replacements in \eqref{eq.sub}, the scalar field \eqref{eq1.7} for the metric \eqref{eq1.21} can be written as follows
\begin{equation}\label{eq.psi}
\psi (r) =\sqrt{\frac{- \nu}{2}} \ln \left(\frac{r - m - \sqrt{m^2 + n^2}}{r - m + \sqrt{m^2 + n^2}}\right)~.
\end{equation}
To find the space-time singularities, one must calculate the Kretschmann or Ricci scalar. The Ricci scalar has a simple form as below
\begin{equation}\label{eq1.222}
\mathcal{R} = -\frac{2 \nu \left(m^2 + n^2\right)}{\left(r^2+ n^2\right)\,(r^2-2\,m\,r-n^2)}\,\Big(1+\frac{(m^2+n^2)\,\sin^2\theta}{r^2-2\,m\,r-n^2}\Big)^{-\nu}.
\end{equation}
It should be noted that the Ricci scalar is not equal to zero in the presence of a massless scalar field. Eq. \eqref{eq.psi} implies that $ \nu $ is always less than or equal to zero, so the singularity of the Ricci scalar for $ \nu\le 0 $ is as follows
\begin{equation}\label{eq1.223}
r_{singularity}=
m+\sqrt{m^2+n^2}.
\end{equation}
Therefore, the curvature singularity in the presence of a scalar field  is the same as that of NUT metric. Eq. \eqref{eq1.223}, represents a naked singularity. As is depicted in Fig. \ref{fig:nutricci}, for constants $m=1$, $n=1$, $\nu=-1$, and different values of $ \theta $, the Ricci scalar diverges for the class of TNS metrics at $ r=1+\sqrt{2} $.

\begin{figure}[H]
\centering
\includegraphics[width=0.7\linewidth]{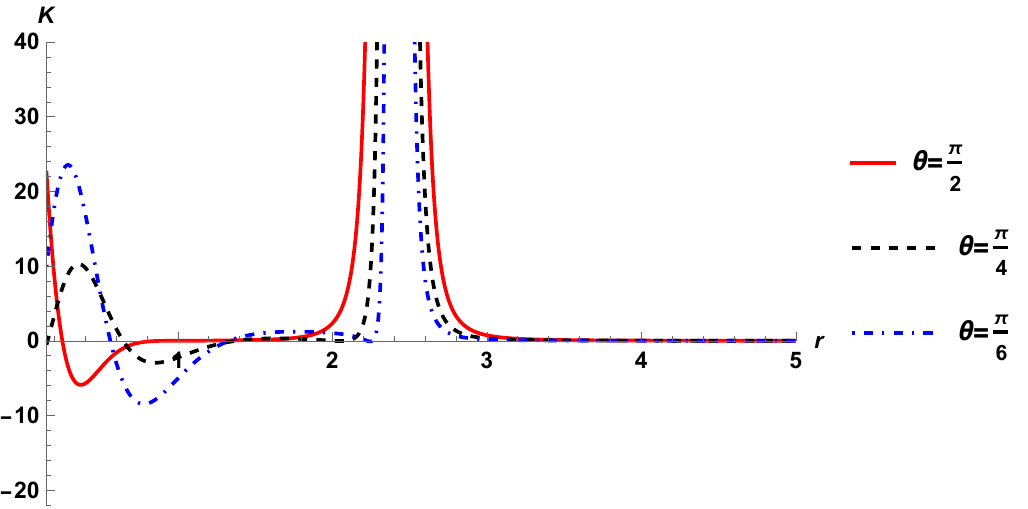}
\caption{The Kretschmann scalar for a TNS metric diverges at $r=1+\sqrt{2}$ for different values of $ \theta $, where $m=1$, $n=1$ and $\nu=-1$.}
\label{Krsh}
\end{figure}

The Kretschmann scalar for the class of TNS metrics in Eq. \eqref{eq1.21}  is as follows
\begin{equation}\label{k1}
K=\frac{4\,X(r,\,\theta)}{(n^2+r^2)^6\,\big(n^2+2\,m\,r-r^2\big)^{2\,(1-\nu)}}.
\end{equation}
where $ X(r,\,\theta) $ is a regular function of $r$ and $\theta$ and represented in Appendix A. According to Eq. \eqref{eq.psi}, $ \nu $ for the class of TNS metrics is always smaller than zero, and therefore, the singular point for the Kretschmann scalar is determined as follows
\begin{equation}\label{k2}
r_{singularity}=
m+\sqrt{m^2+n^2}.
\end{equation}
Therefore, the presence of a scalar field has no effect on the singularity point of metrics with NUT parameter. In Fig. \ref{Krsh}, the Kretschmann scalars are depicted with respect to $r$ for different values of $\theta$. It should be noted that the curvature singularity of the Ricci and Kretschmann scalars occurs at the same point.

\subsection{SINGULARITIES ALONG THE SYMMETRY AXIS $\theta=0$ AND $\theta=\pi$}\label{axis-singularity}
We can rewrite the class of TNS metrics in Eq. \eqref{eq1.21} as below
\begin{equation}\label{t1}
ds^2=-\mathcal{F}(r)\,\big(dt+2\,n\,\cos\theta\,d\phi\big)^2+\frac{\mathcal{K}(r,\,\theta)^\nu}{\mathcal{F}(r)}\,dr^2+(r^2+n^2)\,\big(\mathcal{K}(r,\,\theta)^\nu\,d\theta^2+\sin^2\theta\,d\phi^2\big),
\end{equation}
where
\begin{equation}\label{t2}
\begin{aligned}
&\mathcal{F}(r)=1 - \frac{2 \left(n^2 + m r\right)}{r^2 + n^2},\\
&\mathcal{K}(r,\,\theta)=1+ \frac{\left(m^2 + n^2\right) \sin^2 (\theta)}{r^2 - 2 m r - n^2}.
\end{aligned}
\end{equation}
At $ \nu=0 $, the scalar field becomes zero and metric in Eq. \eqref{t1} becomes NUT metric. When $ r=m+\sqrt{m^2+n^2} $, TNS metrics become singular, and unlike metrics \eqref{eq1.4}, the class of TNS metrics do not have singularity at $r = 0$. Also, metric \eqref{t1} is singular at  $ \theta=0 $ and $ \theta=\pi $ because the signature of the metric changes and is not $ (-,\,+,\,+,\,+) $. This singularity of TNS metrics on the $ \theta=0 $ and $ \theta=\pi $ axes is exactly the same as the Taub-NUT metric. We can regularize the axis at $ \theta=0 $ by using the transformation $ t=\bar{t}-2\,n\,\phi $ for the class of metrics \eqref{t1}, and transform the metric into the following form
\begin{equation}\label{t3}
\begin{aligned}
ds^2=&-\mathcal{F}(r)\,\big(d\bar{t}+2\,n\,(\cos\theta-1)\,d\phi\big)^2+\frac{\mathcal{K}(r,\,\theta)^\nu}{\mathcal{F}(r)}\,dr^2\\
&+(r^2+n^2)\,\big(\mathcal{K}(r,\,\theta)^\nu\,d\theta^2+\sin^2\theta\,d\phi^2\big).
\end{aligned}
\end{equation}
As the axis of symmetry $\theta=0$ is regular in \eqref{t3}, therefore we can consider $\phi$ as a periodic coordinate at $\phi=0$ and $\phi=2\,\pi$. However, at $\theta=\pi$ there is conical singularity as in the NUT metric. Also, by applying the transformation $ t=\bar{t}+2\,n\,\phi $ on metric \eqref{t1}, the metric becomes as follows and the singularity on $ \theta=\pi $ disappears, but we still have a singularity on $ \theta=0 $ axis
\begin{equation}\label{t3b}
\begin{aligned}
ds^2=&-\mathcal{F}(r)\,\big(d\bar{t}+2\,n\,(\cos\theta+1)\,d\phi\big)^2+\frac{\mathcal{K}(r,\,\theta)^\nu}{\mathcal{F}(r)}\,dr^2\\
&+(r^2+n^2)\,\big(\mathcal{K}(r,\,\theta)^\nu\,d\theta^2+\sin^2\theta\,d\phi^2\big).
\end{aligned}
\end{equation}
For the physical description of Taub-NUT metric and its singularities, there are various interpretations, for example, Bonnor \cite{bonnor1969new} considers this singularity as a physical singularity and tries to provide a physical interpretation for it, but Misner tries to remove the singularity \cite{misner1963flatter, misner1967contribution} and for this he uses closed timelike curves (CTC), which will be discussed in the next section. As can be seen from Eqs. \eqref{t3} and \eqref{t3b}, the presence of the scalar field does not affect the singularities at $\theta=0$ and $\theta=\pi$. Therefore, the interpretations that have been used for the singularities at $\theta=0$ and $\theta=\pi$ for the NUT metric are also correct for the class of TNS metrics.
\subsection{THE CLOSED TIMELIKE CURVES IN TNS METRICS}\label{CTC}
In general relativity there are solutions with closed timelike curves (CTC) \cite{tipler1976causality, visser1995lorentzian, rm1984general, lobo2010closed}. If the CTC exists in a spacetime, time travel is possible, and an observer moving along the CTC can go back to an event in the past. This can violate the principle of causality. Also, the tipping over of light cones is one of the consequences of the existence of closed timelike curves. One of the solutions of general relativity, in which the existence of closed timelike curves is possible, is the Taub-NUT metric. In this part, we are going to investigate the effect of the presence of a scalar field on closed time curves in TNS metrics. 

In metrics \eqref{t3} and \eqref{t3b}, a curve with coordinates $ \tau=(t=\text{const},\,r=\text{const},\,\theta=\text{const}) $ has the following invariant length
\begin{equation}\label{t5}
s_\tau^2=(2\,\pi)^2\,\left(- \frac{ 4 n^2 \left(r^2 - 2 m r - n^2\right) \cos^2 (\theta)}{r^2 + n^2} + \left(r^2 + n^2 \right)\sin^2(\theta) \right)=(2\,\pi)^2\,Y(r,\theta).
\end{equation}
The integral curve in $(t=\text{const},\,r=\text{const},\,\theta=\text{const})$ forms a closed timelike curve if $ Y(r,\theta)<0 $ in Eq. \eqref{t5} and forms a closed null curve if $ Y(r,\theta)=0 $. 

As it is clear from the calculations of this part, the presence of the scalar field has no effect on the closed timelike curves and the calculations of the closed timelike curves are the same for the Taub-NUT and TNS metrics.
\subsection{GEODESICS AND TOPOLOGICAL CHARGES}
\begin{figure}[H]
\centering
\includegraphics[width=0.7\linewidth]{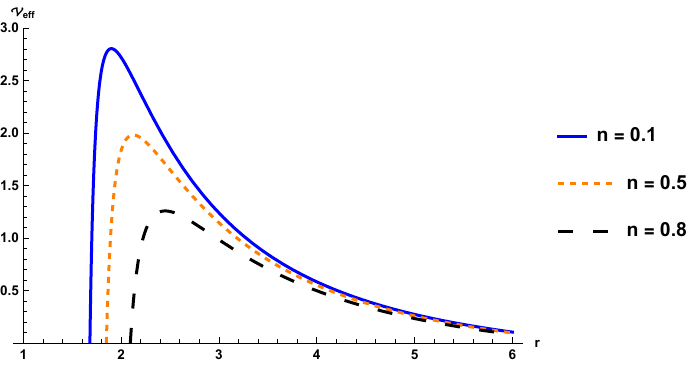}
\caption{The effective potential with respect to r for different values of NUT charge and the constants $m=0.8$, $E=0.5$, $L=4$, and $\kappa=0$.}
\label{fig:veff}
\end{figure}

We use the Lagrangian and the effective potential method to investigate geodesics in the TNS space-time. It should be noted that on the equatorial plane $\theta=\pi/2$, the movement of the test particle is in a circular orbit. Here, the metric components do not depend on $\phi$ and $t$, therefore, two Killing vector fields are defined as follows
\begin{gather}
\xi^{\mu} = \left(\frac{\partial}{\partial t}, 0, 0, 0\right)~,\label{eq1.22}\\
\zeta^{\mu} = \left(0, 0, 0, \frac{\partial}{\partial \phi}\right)~.\label{eq1.23}
\end{gather}
When the observer moves with four-velocity $u^{\mu}$ in the direction of these two vector fields, the metric components will not change, and as a result, we are dealing with two constants here, $\xi^{\mu}$ associated with the constant energy $E$ and $\zeta^{\mu}$ is related to the constant quantity of angular momentum $L$ of the test particles.
\begin{gather}
- E = g_{\mu \nu} u^{\mu} \xi^{\nu} = g_{t t} \dot{t} + g_{t \phi} \dot{\phi}~,\label{eq1.24}\\
L = g_{\mu \nu} u^{\mu} \zeta^{\nu} = g_{t \phi} \dot{t} + g_{\phi \phi} \dot{\phi}~.\label{eq1.25}
\end{gather}
Solving the two Eqs. (\ref{eq1.24}) and (\ref{eq1.25}) simultaneously, we have
\begin{gather}
\dot{t} = \frac{1}{B} \left(E g_{\phi \phi} + L g_{t \phi}\right)~,\label{eq1.26}\\
\dot{\phi} = -\frac{1}{B} \left(E g_{t \phi} + L g_{t t}\right)~,\label{eq1.27}
\end{gather}
where
\begin{equation}
B = g_{t \phi}^{2} - g_{t t} g_{\phi \phi}~. \label{eq1.28}
\end{equation}
Therefore, geodesics related to the coordinates can be obtained by solving Eqs. \eqref{eq1.26} and \eqref{eq1.27}. It is possible to derive the geodesic related to the radial coordinate by defining a Lagrangian as follows
\begin{equation}
\mathcal{L} = \frac{1}{2} g_{\mu \nu} \dot{x}^{\mu} \dot{x}^{\nu} = - \frac{1}{2} \kappa~, \label{eq1.29}
\end{equation}
where $\kappa=1, 0, -1$ are affine parameters for time-like, light-like, and space-like paths, respectively. Also, conjugate momentum is defined as
\begin{equation}
\pi_{\mu} = \frac{\partial \mathcal{L}}{\partial \dot{x}^{\mu}} = g_{\mu \nu} \dot{x}^{\nu}~. \label{eq1.30}
\end{equation}
The Hamiltonian for the test particle for the TNS metric is
\begin{equation}
\mathcal{H} = \pi_{\mu} \dot{x}^{\mu} - \mathcal{L} = \frac{1}{2} \left(g_{t t} \dot{t}^2 + g_{r r} \dot{r}^2 + g_{\theta \theta} \dot{\theta}^2 + g_{\phi \phi} \dot{\phi}^2 + 2 g_{t \phi} \dot{t} \dot{\phi}\right) = - \frac{1}{2} \kappa~, \label{eq1.31}
\end{equation}
or
\begin{equation}
g_{t t} \dot{t}^2 + g_{r r} \dot{r}^2 + g_{\theta \theta} \dot{\theta}^2 + g_{\phi \phi} \dot{\phi}^2 + 2 g_{t \phi} \dot{t} \dot{\phi} + \kappa = 0~. \label{eq1.32}
\end{equation}
The kinetic energy $K$ and potential energy $\mathcal{V}$ for particles are defined as below
\begin{gather}
K = g_{r r} \dot{r}^2 + g_{\theta \theta} \dot{\theta}^2~,\label{eq1.33}\\
\mathcal{V} = g_{t t} \dot{t}^2 + g_{\phi \phi} \dot{\phi}^2 + 2 g_{t \phi} \dot{t} \dot{\phi} + \kappa~.\label{eq1.34}
\end{gather}
Therefore, the trajectory of the test particle corresponds to the following relation
\begin{equation}
K + \mathcal{V} = 0~. \label{eq1.35}
\end{equation}
From the Eqs. \eqref{eq1.26}, \eqref{eq1.27}, and \eqref{eq1.34}, the effective potential can be written as
\begin{equation}
\mathcal{V}_{eff} = -\frac{1}{B} \left(E^2 g_{\phi \phi} + 2 E L g_{t \phi} + L^2 g_{t t}\right) + \kappa~. \label{eq1.36}
\end{equation}
Therefore, by using metric in Eq. \eqref{eq1.21}, we can find the effective potential for the TNS metric. In Fig. \ref{fig:veff}, the effective potentials are depicted for different values of the NUT charge and $m=0.8$, $E=0.5$, $L=4$, and $\kappa=0$.

The stable bound orbits for TNS metric can be found using the conditions
\begin{equation}\label{eq1.361}
\partial^{2}_{r} \mathcal{V}_{eff} \partial^{2}_{\theta} \mathcal{V}_{eff} - \left(\partial_{r} \partial_{\theta} \mathcal{V}_{eff}\right)^2 \textgreater 0, \hspace{10pt} \frac{1}{g_{r r}} \partial^{2}_{r} \mathcal{V}_{eff} + \frac{1}{g_{\theta \theta}} \partial^{2}_{\theta} \mathcal{V}_{eff} \textgreater 0~. 
\end{equation}
Since we examine the light rings, we must deal with the ones in which $\kappa=0$. As a result, the effective potential can be written as
\begin{equation}\label{eq1.37}
\mathcal{V}_{eff} = - \frac{g_{\phi \phi}}{B} \left(E - E_{1}\right) \left(E - E_{2}\right)~,
\end{equation}
where
\begin{gather}
E_{1} = \frac{-L g_{t \phi} + L \sqrt{B}}{g_{\phi \phi}}~,\label{eq1.38}\\
E_{2} = \frac{-L g_{t \phi} - L \sqrt{B}}{g_{\phi \phi}}~.\label{eq1.39}
\end{gather}
Considering that the time-like circular orbits for a test particle satisfies $\mathcal{V}_{eff} = 0$ and $\partial_{r} \mathcal{V}_{eff} = 0$, using Eqs. \eqref{eq1.37} and \eqref{eq1.38}, light ring radius $r_{LR}$ obtains as \cite{Cunha2016, Ye2023, Cunha2020,Wei2020, Guo2021}
\begin{equation}\label{eq1.44}
\Phi^r = \frac{\partial_{r} E_{1}}{\sqrt{g_{r r}}} = 0~, \hspace{10pt} \Phi^{\theta} = \frac{\partial_{\theta} E_{1}}{\sqrt{g_{\theta \theta}}} = 0~,
\end{equation}
where ($r$, $\theta$) represent the light ring coordinates. Using $E_{1}$ in the two dimensional space-time $\left(r, \theta \right)$ we introduce a vector field as
\begin{equation}\label{eq1.43}
\Phi^r = \frac{\partial_{r} E_{1}}{\sqrt{g_{r r}}}~, \hspace{10pt} \Phi^{\theta} = \frac{\partial_{\theta} E_{1}}{\sqrt{g_{\theta \theta}}}~.
\end{equation}
As a result, from the topological current point of view, these points form a topological number so that a topological current can be attributed to it. In the following, by introducing the topological current and the topological number, the vector field will be constructed for the TNS space-time, and the light ring will be determined using zero points of the vector field.

It is known that light rings around black holes can be studied by using   topological charges corresponding to them \cite{duan1979TSU, duan1984structure,Duan2000, Cunha2020, Wei2020, Guo2021, Wei2022, Duan1984, Cunha2017}. The topological current is defined as
\begin{equation}\label{eq1.45}
j^{\mu} = \frac{1}{2 \pi}\, \epsilon^{\mu \nu \rho}\, \epsilon_{a b}\, \frac{\partial n^a}{\partial x^\nu}\, \frac{\partial n^b}{\partial x^\rho}.
\end{equation}
where $ n^a=(\frac{\phi^r}{\lvert \phi \rvert},\,\frac{\phi^\theta}{\lvert \phi \rvert}) $ is a unit vector in the direction of $\phi$ and $ x^\mu=(t,\,r,\,\theta) $. We can simply show that $j^{\mu}$ is a conserved current, that is, we have
\begin{gather}
\partial_{\mu} j^{\mu} = 0~.\label{eq.46}
\end{gather}
As a result, a constant quantity can be related to this conserved current, which is the topological charge. Using the following integral over the parameter region $\Sigma$, the constant charge is obtained as
\begin{equation}\label{eq1.47}
Q = \int_{\Sigma} j^0 d^2 x~.
\end{equation}
Since this current is caused by point-like topological charges that are placed on the light ring, it can be obtained using the definition of Dirac's delta function as
\begin{equation}\label{eq1.48}
j^{\mu} = \int_{\Sigma} \delta^2 \left(\Phi\right) j^\mu \left(\frac{\Phi}{x}\right)~.
\end{equation}
This equation is completely in agreement with the concept of topological current because the topological current is non-zero only on the light ring. At the topological charge points, the vector field $\Phi$ is $0$, and according to the definition, the Dirac's delta function \eqref{eq1.48} is non-zero, as a result, we have topological current. It should be noted that $\Phi = \left(X^i, t\right)$ are the vector field components obtained from the effective potential. As a result, the topological charge is simplified as
\begin{equation}\label{eq1.49}
Q = \int_{S} j^{0} d^2 x = \sum_{i=1}^{N} \beta_{i} \eta_{i} = \sum_{i=1}^{N} w_{i}~,
\end{equation}
in which $\beta_{i}$ is the Hopf index, $\eta= \pm 1$ is the Brouwer degree, and the $w_{i}$ is the winding index for the zero points of the vector field $\Phi^\alpha$ in $ S $ area. In Eq. \eqref{eq1.49}, at the points where the vector field $\Phi$ is zero, the topological charge is non-zero. Note that these points are located on the light ring so that each light ring will correspond to one topological charge. If $ S $ covers zero points of $\Phi$, then topological charge $Q$ corresponds to the sum of the winding index of these points on the light ring. On the other hand, if only one zero point of the vector field $\Phi$ is covered by the $ S $, then the topological charge $Q$ will be directly proportional to the winding number.

\begin{figure}[H]
\centering
\includegraphics[width=0.7\linewidth]{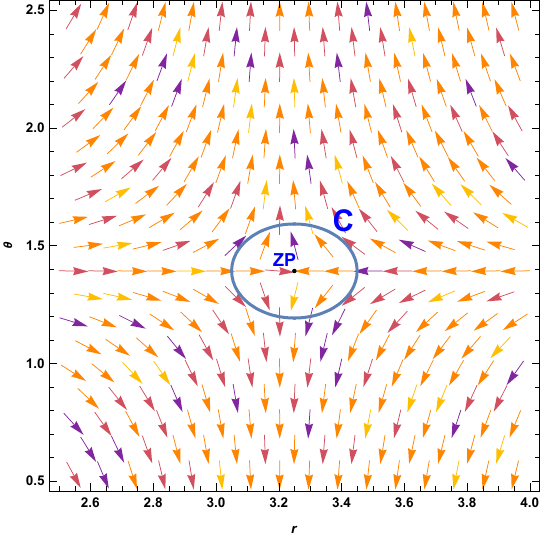}
\caption{The unit vector field for $m=1$, $\mu=0$, $l=1.9$, $\nu=-1$, and $n=0.5$.}
\label{fig:topcha}
\end{figure}

As it can be seen in Fig. \ref{fig:topcha}, for specific values of the parameters of the TNS metric, the topological charge $Q$ is $-1$, which indicates that there is an unstable orbit for massless particles in TNS space-time.

\section{NEW CLASS OF EXACT SOLUTIONS OF THE EINSTEIN-CONFORMAL-SCALAR THEORY}\label{sec:Cthree}
Bekenstein has shown that the Einstein-scalar theory and the Einstein-conformal-scalar theory are coformally related and obtained some of its exact solutions in \cite{Bekenstein1947}, which is the conformal transformation of the FJNW metric. In the following, we take the three-parameter metric in Eq. \eqref{eq1.4} as a seed metric and by introducing a conformal transformation derive a new class of exact solutions of the Einstein-conformal-scalar theory.\\
Consider the following action of the Einstein-conformal-scalar theory that is related to the action of the Einstein-scalar theory in Eq. \eqref{eq1.1}, by a conformal transformation as follows
\begin{equation}\label{eq3.1}
S = \int d^4 x \sqrt{- \bar{g}} \left(\bar{\mathcal{R}} - \bar{\nabla}_{\mu} \bar{\psi} \bar{\nabla}^{\mu} \bar{\psi} - \frac{1}{6} \bar{\mathcal{R}} \bar{\psi}^2\right)~,
\end{equation}
where
\begin{gather}
\bar{g}_{\mu \nu} = \cosh^2 \left(\sqrt{\frac{1}{6}} \psi \right) g_{\mu \nu}~, \label{eq3.7} \\
\bar{\psi} = \sqrt{6} \tanh \left(\sqrt{\frac{1}{6}} \psi\right)~. \label{eq3.8}
\end{gather}
Varying the action \eqref{eq3.1} with respect to scalar field $\bar{\psi}$, and spacetime metric we derive the following field equations
\begin{gather}
\left(\bar{\nabla}_{\mu} \bar{\nabla}^{\mu} - \frac{1}{6} \bar{\mathcal{R}}\right) \bar{\psi} = 0~, \label{eq3.2} \\
\bar{G}_{\mu \nu} = \bar{T}_{\mu \nu}~, \label{eq3.3}
\end{gather}
where $\bar{G}_{\mu \nu}$ is the conformal Einstein tensor and the energy-momentum tensor is
\begin{equation}\label{eq3.4}
\bar{T}_{\mu \nu} = \bar{\nabla}_{\mu} \bar{\psi} \bar{\nabla}_{\nu} \bar{\psi} - \frac{1}{2} \bar{g}_{\mu \nu} \bar{\nabla}_{\mu} \bar{\psi} \bar{\nabla}^{\mu} + \frac{1}{6} \left(\bar{g}_{\mu \nu} \bar{\nabla}_{\mu} \bar{\nabla}^{\mu} - \bar{\nabla}_{\mu} \bar{\nabla}_{\nu} + \bar{G}_{\mu \nu}\right)\bar{\psi}^2~.
\end{equation}
The new solution should solve the transformed field equations. It is straightforward to show that for a conformal non-minimal coupling, the Ricci scalar vanishes ($\bar{\mathcal{R}} = 0$) and the equation of motion for $\bar{\psi}$ in Eq. \eqref{eq3.2} reduces to
\begin{equation}\label{eq3.6}
\bar{\nabla}_{\mu} \bar{\nabla}^{\mu} \bar{\psi} = 0~.
\end{equation}
We can obtain exact class of solutions that satisfies the field Eqs. \eqref{eq3.2} and \eqref{eq3.3} by using conformal transformations \eqref{eq3.7} and \eqref{eq3.8} of the class of three-parameter \eqref{eq1.4} and TNS metrics \eqref{eq1.21}. Using the two transformations in Eqs. \eqref{eq3.7} and \eqref{eq3.8} with the scalar field (\ref{eq1.7}) and replacing $m$ by $2\, M$, and using the coordinates $(t, \tilde{r}, \theta, \phi)$ with
\begin{equation}
r = \frac{\tilde{r}^2}{\tilde{r}-M}~,\label{eq3.9} \\
\end{equation}
and using Eqs. \eqref{eq1.4}, \eqref{eq1.7}, the rescaled metric $\bar{g}_{\mu \nu}$ and the scalar field $\bar{\psi}$ are obtained as follows
\begin{gather}
\bar{g}_{\mu \nu} = \frac{\left(\left(1- \frac{2 M}{\tilde{r}}\right)^{2 \sqrt{\frac{1-\gamma^2 - \nu}{3}}} + 1\right)^2}{4 \left(1-\frac{2 M}{\tilde{r}}\right)^{2 \sqrt{\frac{1-\gamma^2 - \nu}{3}}}} g_{\mu \nu}~, \label{eq3.10} \\
\bar{\psi} = \sqrt{6} \frac{\left(1-\frac{2 M}{ \tilde{r}}\right)^{2 \sqrt{\frac{1-\gamma^2 - \nu}{3}}} -1}{\left(1- \frac{2 M}{\tilde{r}}\right)^{2 \sqrt{\frac{1-\gamma^2 - \nu}{3}}} +1}~. \label{eq3.11}
\end{gather}
Therefore, the new exact class of three parameter solutions of equation \eqref{eq3.3} can be written as follows
\begin{equation}\label{eq3.12}
\begin{aligned}
ds^2 = \frac{1}{4} & \left(\left(1 -\frac{2 M}{\tilde{r}}\right)^{-\sqrt{\frac{\gamma-\gamma^2 + \mu}{3}}} + \left(1-\frac{2 M}{\tilde{r}}\right)^{\sqrt{\frac{\gamma-\gamma^2 + \mu}{3}}}\right)^{2} \biggl[-\left(1-\frac{2 M}{\tilde{r}}\right)^{2 \gamma} dt^2 + \\
&\frac{\left(1-\frac{2 M}{\tilde{r}}\right)^{2 \mu} \tilde{k}^{1-\gamma-\mu}}{\left(1-\frac{M}{\tilde{r}}\right)^{2}} \left(\frac{dr^2}{\left(1-\frac{M}{\tilde{r}}\right)^2} + \tilde{r}^{2} d\theta^2 \right) + \frac{\tilde{r}^{2} \left(1-\frac{2 M}{\tilde{r}}\right)^{2-2\gamma}}{\left(1-\frac{M}{\tilde{r}}\right)^2} \sin^2(\theta) d\phi^2 \biggl]~,
\end{aligned}
\end{equation}
where
\begin{equation}
\tilde{k} = \frac{\tilde{r}^2 \left(1-\frac{2 M}{\tilde{r}}\right)^2 + 4 M^2 \left(1-\frac{M}{\tilde{r}}\right)^2 \sin^2 (\theta)}{\tilde{r^2}}. \label{eq3.13}
\end{equation}
It should be noted that the metric in Eq. \eqref{eq3.12} represents a class of exact solutions of the Einstein-conformal-scalar theory with three parameters $M$, $\gamma$, and $\mu$ with the condition $\mu+\nu=1-\gamma$. A special case is when $\gamma=1$ and $\mu= -\nu$, which can be written as follows
\begin{equation}\label{eq3.14}
\begin{aligned}
ds^2& = \frac{1}{4}\left(\left(1-\frac{2 M}{\tilde{r}}\right)^{-\sqrt{\frac{\mu}{3}}} + \left(1-\frac{2 M}{\tilde{r}}\right)^{\sqrt{\frac{\mu}{3}}}\right)^{2} \biggl[-\left(1-\frac{2 M}{\tilde{r}}\right)^{2} dt^2\\ &+\frac{\left(1-\frac{2 M}{\tilde{r}}\right)^{2 \mu} \tilde{k}^{-\mu}}{\left(1-\frac{M}{\tilde{r}}\right)^{2}} \biggl(\frac{dr^2}{\left(1-\frac{M}{\tilde{r}}\right)^2} +\tilde{r}^{2} d\theta^2 \biggl) + \frac{\tilde{r}^{2}}{\left(1-\frac{M}{\tilde{r}}\right)^2} \sin^2(\theta) d\phi^2 \biggl]~.
\end{aligned}
\end{equation}
For certain circular paths $r_{0}$ located in the plane $\theta=\frac{\pi}{2}$ considering the conformal metric (\ref{eq3.14}) and using the effective potential in \eqref{eq1.34}, we have
\begin{gather}
\bar{g}_{t t} + \bar{g}_{\phi \phi} \left(\frac{d \phi}{dt}\right)^{2} = 0 ~, \label{eq3.15} \\
\bar{\Gamma}^{r}_{\hspace{3pt} t t} + \bar{\Gamma}^{r}_{\hspace{3pt} \phi \phi} \left(\frac{d\phi}{dt}\right)^{2} = 0~.\label{eq3.16}
\end{gather}
By substituting the components of the conformal metric \eqref{eq3.14} and the related Christoffel symbols into Eqs. \eqref{eq3.15} and \eqref{eq3.16}, one obtain the following relations
\begin{gather}
\frac{d\phi}{dt} = \frac{\left(1-\frac{2 M}{\tilde{r}}\right) \left(1-\frac{M}{\tilde{r}}\right)}{\tilde{r}} ~, \label{eq3.17} \\
\frac{d\phi}{dt} = \sqrt{- \frac{\bar{\Gamma}^{r}_{\hspace{3pt} t t}}{\bar{\Gamma}^{r}_{\hspace{3pt} \phi \phi}}}~.\label{eq3.18}
\end{gather}
Now, as the right sides of Eqs. \eqref{eq3.17} and \eqref{eq3.18}, are equal, the radius of the light ring can be obtained as
\begin{equation}\label{eq3.111}
\tilde{r} = \left(3 + \sqrt{3}\right) M
\end{equation}
\section{CONFORMAL-TAUB-NUT-SCALAR SOLUTION}\label{sec:CTNS}
In section \ref{sec:TNS}, we used Ernst potential and Ehlers transformations to derive the TNS metric \eqref{eq1.21} in Einstein-Scalar theory. Now, we tend to find the Conformal-Taub-NUT-Scalar (CTNS) metric using the conformal transformations \eqref{eq3.7} and \eqref{eq3.8} that is a class of exact solutions ($\gamma=1$, $\mu= -\nu$)  of Eq. \eqref{eq3.3}. For this, we replace $m\rightarrow 2\,M$ and $n\rightarrow 2\,N$ and use the coordinates $(t, \tilde{r}, \theta, \phi)$ with
\begin{gather}
r = \frac{\tilde{r}^2}{\tilde{r} - M}~,\label{eq3.19} \\
\bar{\rho}^2 = M^2 + N^2~, \label{eq3.20} \\
\bar{\Delta} = \tilde{r}^2 - 2 M \tilde{r} + M^2~.\label{eq3.21}
\end{gather}
Using Bekenstein's approach and considering the scalar field \eqref{eq.psi} and the metric \eqref{eq1.21} with transformations \eqref{eq3.10} and \eqref{eq3.11}, the conformal scalar field and conformal metric can be obtained as follows
\begin{gather}
\bar{g}_{\mu \nu} = \cosh^2 \left(\sqrt{-\frac{\nu}{12}} \ln \left( \frac{\bar{\Delta} - 2 \bar{\rho} \left(\tilde{r} - M\right) + M^2}{\bar{\Delta} + 2 \bar{\rho} \left(\tilde{r} - M\right) + M^2}\right)\right)  g_{\mu \nu}~, \label{eq3.22} \\
\bar{\psi} =\sqrt{6} \frac{\left(\bar{\Delta} - 2 \bar{\rho} \left(\tilde{r} - M\right) + M^2\right)^{\sqrt{-\frac{\nu}{3}}} - \left(\bar{\Delta} + 2 \bar{\rho} \left(\tilde{r} - M\right) + M^2\right)^{\sqrt{-\frac{\nu}{3}}}}{\left(\bar{\Delta} - 2 \bar{\rho} \left(\tilde{r} - M\right) + M^2\right)^{\sqrt{-\frac{\nu}{3}}} + \left(\bar{\Delta} + 2 \bar{\rho} \left(\tilde{r} - M\right) + M^2\right)^{\sqrt{-\frac{\nu}{3}}}} ~, \label{eq3.23}
\end{gather}
so the CTNS metric components can be derived as follows
\begin{equation}\label{eq3.24}
\begin{aligned}
&\bar{g}_{t t} = - \frac{W - V^2 U}{4 \left(1+ V^2 U\right)} H,\\
&\bar{g}_{r r} =\frac{W^{-1} \left(1+ V^2 U\right)}{4 \left(W - V^2 U\right)} \left(1+ \frac{4 \bar{\rho}^2 \sin^2 (\theta)}{\tilde{r}^2 \left(U^{-1} W - V^2 \right)}\right)^{-\mu} H,\\
&\bar{g}_{\theta \theta} = \frac{\tilde{r}^2}{4} \left(U^{-1} + V^2 \right) \left(1+ \frac{4 \bar{\rho}^2 \sin^2 (\theta)}{\tilde{r}^2 \left(U^{-1} W - V^2 \right)}\right)^{-\mu} H,\\
&\bar{g}_{\phi \phi} = \frac{1}{4} \Big[\frac{16 N^2 \left[(1-V) - \frac{2 M}{\tilde{r}} (1-\frac{N}{\tilde{r}})\right] \left[\frac{2 M}{\tilde{r}} (1+ \frac{N}{\tilde{r}})- (1+ V)\right] \cos^2(\theta)}{1+ V^2 U}\\
&\qquad\ + \tilde{r}^2 \left(U^{-1} + V^2 \right) \sin^2 (\theta)\Big] H,\\
&\bar{g}_{t \phi} = \frac{N \left[(1-V)-\frac{2 M}{\tilde{r}}(1- \frac{N}{\tilde{r}})\right] \left[\frac{2 M}{\tilde{r}}(1+\frac{N}{\tilde{r}})-(1+V)\right]}{1+V^2 U} H \cos(\theta),
\end{aligned}
\end{equation}
were $H$, $W$, $U$, and $V$ are
\begin{equation}\label{eq3.25}
\begin{gathered}
H = \left( \frac{\bar{\Delta} - 2 \bar{\rho} \left(\tilde{r} - M\right) + M^2}{\bar{\Delta} + 2 \bar{\rho} \left(\tilde{r} - M\right) + M^2}\right)^{-\sqrt{\frac{\mu}{3}}} \left(1+ \left( \frac{\bar{\Delta} - 2 \bar{\rho} \left(\tilde{r} - M\right) + M^2}{\bar{\Delta} + 2 \bar{\rho} \left(\tilde{r} - M\right) + M^2}\right)^{\sqrt{\frac{\mu}{3}}} \right)^2~, \\
W = \left(1-\frac{2 M}{\tilde{r}}\right)^2~, \\
U = \left(1- \frac{M}{\tilde{r}}\right)^2~, \\
V = \frac{2 N}{\tilde{r}}~,
\end{gathered}
\end{equation}
which satisfies the field equation in \eqref{eq3.3}. The Ricci scalar for metric in \eqref{eq3.24} is too complicated to be written here, however, we have derived  its singularities as follows
\begin{equation}\label{eq3.26}
\begin{gathered}
\tilde{r} = M + N \pm \sqrt{M^2 + N^2}~, \\
\tilde{r} = M - N + \sqrt{M^2 + N^2}~.
\end{gathered}
\end{equation}
It should be noted that the conformal transformation dose not remove the naked singularities of the TNS metric in Eq. \eqref{h180_label}. However, the curvature singularities of the CTNS metric differs from the singularities of TNS metric in Eq. \eqref{eq1.223}.
\subsection{GEODESICS IN CTNS SPACE-TIME}
\begin{figure}[H]
\centering
\includegraphics[width=0.7\linewidth]{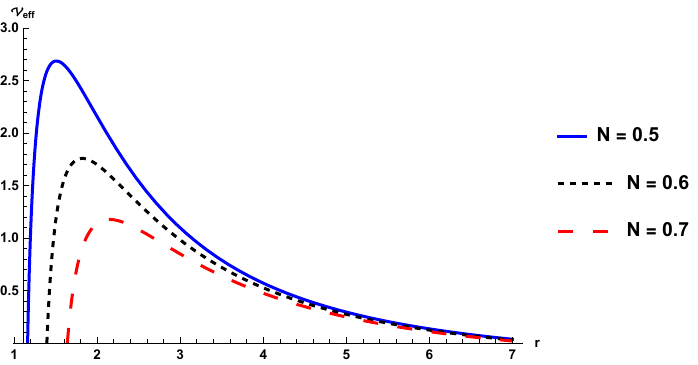}
\caption{The effective potential for CTNS metric \eqref{eq3.24} with respect to r for different values of NUT charge and the constants $M=0.1$, $E=0.5$, $L=4$, $\nu = -1$, and $\kappa=0$.}
\label{fig:ctnsveff}
\end{figure}

In order to investigate the geometric features of the CTNS space-time, we determine the equations of motion of test particles in this space-time. Considering the CTNS metric \eqref{eq3.24}, the geodesic equation is written as
\begin{equation}\label{eq33}
\frac{d^2 x^{\mu}}{ds^2} + \bar{\Gamma}^{\mu}_{\hspace{3pt} \nu \rho} \frac{dx^\nu}{ds} \frac{dx^\rho}{ds} = 0~,
\end{equation}
where $x^\mu = x^\mu (s)$ is the test particle paths, $s$ is the affine parameter, and $\bar{\Gamma}^{\mu}_{\hspace{3pt} \nu \rho}$ is the Christoffel symbol for the CTNS metric $\bar{g}_{\mu \nu} = \Omega(x) g_{\mu \nu}$ whic is defined as
\begin{equation}\label{eq34}
\bar{\Gamma}^{\mu}_{\hspace{3pt} \nu \rho} = \Gamma^{\mu}_{\hspace{3pt} \nu \rho} + \frac{1}{2} \left(\delta^{\mu}_{\nu} \partial_{\rho} \ln \Omega + \delta^{\mu}_{\rho} \partial_{\nu} \ln \Omega - g_{\nu \rho} g^{\mu \lambda} \partial_{\lambda} \ln \Omega\right)~,
\end{equation}
where
\begin{equation}\label{eq35}
\Omega(x) = \Omega(\tilde{r}, \theta) = \cosh^2 \left(\sqrt{-\frac{\nu}{12}} \ln \left( \frac{\bar{\Delta} - 2 \bar{\rho} \left(\tilde{r} - M\right) + M^2}{\bar{\Delta} + 2 \bar{\rho} \left(\tilde{r} - M\right) + M^2}\right)\right)~.
\end{equation}
Substituting Eq. \eqref{eq34} in Eq. \eqref{eq33} and using $d\bar{s} = \Omega^{-1} ds$ we have
\begin{equation}\label{eq36}
\frac{d^2 x^\mu}{d\bar{s}^2} + \Gamma^{\mu}_{\hspace{3pt} \nu \rho} \frac{dx^\nu}{d\bar{s}}\frac{dx^\rho}{d\bar{s}} = F^{\mu}~,
\end{equation}
where $F^{\mu} = \frac{\kappa}{2} g^{\mu \lambda} \partial_{\lambda} \ln \Omega$ is the external effective force due to the CTNS metric and $\kappa = g_{\nu \rho} \frac{dx^\nu}{d\bar{s}} \frac{dx^\rho}{d\bar{s}}$. To investigate the time-like orbits in CTNS space-time, we will use the following effective potential
\begin{equation}\label{eq37}
\mathcal{V} (\bar{r}, \theta) = 1+ \frac{L^2 \bar{g}_{t t} + 2 E L \bar{g}_{t \phi} + E^2 \bar{g}_{\phi \phi}}{\bar{g}_{t t} \bar{g}_{\phi \phi} - \bar{g}_{t \phi}^{2}}~.
\end{equation}
In CTNS space-time, $E$ and $L$ are defined as
\begin{gather}
E = - \frac{\bar{g}_{t t} + \bar{g}_{t \phi} \omega}{\sqrt{-\left(\bar{g}_{t t} + 2 \bar{g}_{t \phi} \omega + \bar{g}_{\phi \phi} \omega^2\right)}}~, \label{eq38} \\
L= \frac{\bar{g}_{t \phi} + \bar{g}_{\phi \phi} \omega}{\sqrt{-\left(\bar{g}_{t t} + 2 \bar{g}_{t \phi} \omega + \bar{g}_{\phi \phi} \omega^2\right)}}~, \label{eq39}
\end{gather}
where $\omega=\frac{d \phi}{dt}$ is the angular velocity. Now, it is a matter of calculation to find the effective potential for the CTNS metric. Fig. \ref{fig:ctnsveff} represents the effective potential for CTNS metric for different values of NUT charge and $M=0.1$, $E=0.5$, $L=4$, $\nu = -1$, and $\kappa=0$.

\begin{figure}[H]
\centering
\includegraphics[width=0.7\linewidth]{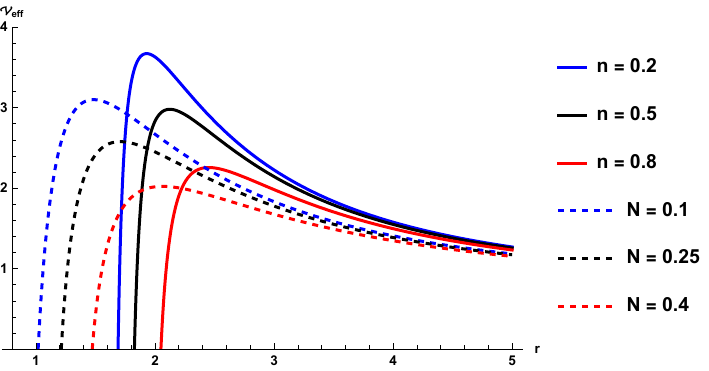}
\caption{The effective potentials for TNS and CTNS metrics for different values of NUT charge and constant values $E=0.5$, $L=4$, $\nu = -1$, $m=0.8$, and $M=0.4$. The straight and dashed lines correspond to TNS and CTNS space-times, respectively.}
\label{fig:comparisonveff}
\end{figure}

Here, we would like to compare the time-like geodesics of TNS and CTNS space-times for $\kappa=1$. For this purpose, we find the effective potential for these two metrics. Fig. \ref{fig:comparisonveff} is the comparison diagram for $E=0.5$, $L=4$, $\nu=-1$, and $\kappa=1$. We have used $m \rightarrow 2\, M$ and $n \rightarrow 2\, N$ to compare the metrics, so each color belongs to a specific value for mass and NUT charge.

\section{QNMs OF TNS AND CTNS METRICS}\label{sec:QNM}
In this section, we will  derive the QNMs of TNS and CTNS metrics using the light ring method in the eikonal limit
\cite{ferrari1984oscillations, ferrari1984new, mashhoon1985stability}.
\subsection{QNMs OF TNS METRIC}\label{62}
The QNMs of the TNS metric can be obtained by assuming $\theta=\pi/2$ in metric \eqref{eq1.21} and then expand the metric components up to the second order of $ n $ and the first order of $\nu$ and omit the terms $n\, \nu$ (we use these approximations until the end of the QNMs calculations)
\begin{equation}\label{50-label}
\begin{aligned}
ds^2=&-\big(1-\frac{2\,m}{r}+\frac{2\,n^2\,(m-r)}{r^3}\big)\,dt^2+\frac{1}{r^3\,(1-2\,m/r)^2}\big[r^3+2\,r\,n^2-2\,m\,r\,(r\\
&+n^2)+r^3\,\nu\,(1-2\,m/r)\,\ln\big(1+\frac{m^2}{r^2\,(1-2\,m/r)}\big)\big]\,dr^2+\Big(n^2+r^2\,\big[1+\nu\,\ln\big(1\\
&+\frac{m^2}{r^2\,(1-2\,m/r)}\big)\big]\Big)\,d\theta^2+(n^2+r^2)\,d\phi^2.
\end{aligned}
\end{equation}
The QNMs are represented by $ Q = \Omega + i \, \Gamma $, which in the light ring method and in the eikonal limit, $ \Omega $ is defined by the following equation
\begin{equation}\label{51-label}
\Omega=\pm\,j\,\Omega_{\pm},
\end{equation}
where $j$ is the member of the integers and $ \Omega_{\pm} $ is the frequency of the circular orbits of the null rays. Also, wave perturbations can be considered using the superposition of eigenmodes as follows
\begin{equation}\label{52-label}
e^{i \, (\Omega \, t - \ell \, \phi)} \, S_{\Omega \, j \, \ell \, s} (r, \theta),
\end{equation}
where, $ S $ and $ \Omega $ are the spin and wave's frequency respectively. Also, $ \ell $ and $ j $ are angular momentum with the following condition
\begin{equation}\label{53-label}
\lvert \ell \rvert \le j.
\end{equation} 
We assume that $ \Omega \gg 1/M $ and $ \lvert \ell \rvert = j \gg 1 $ in the eikonal limit. Moreover, from the numerical calculations of black holes QNMs
\cite{ferrari1984oscillations, ferrari1984new, mashhoon1985stability}
oscillation frequency can be written as follow
\begin{equation}\label{53a-label}
\Omega = \ell \, \frac{d \phi}{d t} = \pm \, j \, \Omega_{\pm},
\end{equation} 
which is equal to the QNMs oscillations. To calculate QNMs, we derive the radius of null circular ($\theta=\pi/2$). In every light ring, we have
\begin{equation}\label{54-label}
g_{t t} + g_{\phi \phi} \, \big( \frac{d \phi}{d t} \big)^2 = 0.
\end{equation}
The radial component of the geodesics is as follows
\begin{equation}\label{55-label}
\Gamma_{t t}^r + \Gamma_{\phi \phi}^r  \, \big( \frac{d \phi}{d t} \big)^2 = 0.
\end{equation}
By placing the metric components \eqref{50-label} in Eqs. \eqref{54-label} and \eqref{55-label}, we have
\begin{equation}\label{56-label}
\begin{aligned}
\frac{d\phi}{dt}=\frac{2\,r^3-3\,r\,n^2+4\,m\,(n^2-r^2)}{2\,r^4\,\sqrt{1-2\,m/r}},
\end{aligned}
\end{equation}
\begin{equation}\label{57-label}
\begin{aligned}
\frac{d\phi}{dt}=\frac{2\,m\,r^2-n^2\,(3\,m-2\,r)}{2\,r^3\,\sqrt{m\,r}}.
\end{aligned}
\end{equation}
By equating Eqs. \eqref{56-label} and \eqref{57-label} and assuming the following perturbation
\begin{equation}\label{58-label}
r = 3 \, m + \bar{\epsilon},
\end{equation}
where $ \bar{\epsilon} $ is a very small parameter that indicates a perturbation, and $ 3 m $ is the light ring radius for the Schwarzschild black hole and $r \equiv r_{0_{TNS}}$. By keeping the perturbation terms up to first order of $ \bar{\epsilon} $, one has 
\begin{equation}\label{59-label}
\bar{\epsilon}=\frac{8\,n^2}{9\,m},\quad \Rightarrow \quad r_{0_{TNS}}=3\,m+\frac{8\,n^2}{9\,m}.
\end{equation}
By placing $ r_{0_{TNS}} $ in Eq. \eqref{56-label} or Eq. \eqref{57-label}, we have
\begin{equation}\label{60-label}
\Omega_\pm=\frac{d\phi}{dt}=\frac{1}{3\,\sqrt{3}\,m}\,\big(1\pm\frac{5}{18}\,(\frac{n}{m})^2\big).
\end{equation} 
Now we have to calculate the other term of QNMs ($ \Gamma $). $\Gamma$ represents the decay rate in the amplitude of QNMs. To calculate $ \Gamma $, we perturb the null equatorial circular orbit. On the other hand, we assume a perturbation in coordinates $ \bar{x}^\mu = (t, r, \theta, \phi) = (t, r_0, \frac{\pi}{2}, \Omega_\pm t) $ as follows
\begin{equation}\label{61-label}
r = r_0 \, [1 + \epsilon \, \delta_r(t)]~, \quad \phi = \Omega_\pm \, [t + \epsilon \, \delta_\phi(t)]~, \quad \ell = t + \epsilon \, \delta_\ell(t),
\end{equation}
where, $ \epsilon $ is perturbation parameter and following conditions for $ \delta_r(t) $, $ \delta_\phi(t) $ and $ \delta_\ell(t) $ are applied
\begin{equation}\label{62-label}
\delta_r(0) = \delta_\phi(0) = \delta_\ell(0) = 0,
\end{equation}
from Eqs. \eqref{53a-label} and \eqref{61-label}, we have $ \delta_\phi(t) = 0 $. The propagation vector is defined as follows (up to $ \mathcal{O}(\epsilon) $ ) 
\begin{equation}\label{63-label}
K^\mu = \frac{d x^\mu}{d \ell} = \big(1 - \epsilon \, \delta_\ell^\prime, \, \epsilon \, r_0 \, \delta_r^\prime, \, 0, \, \Omega_\pm \, (1 - \epsilon \, \delta_\ell^\prime)\big).
\end{equation}
Here, the prime represents the derivative with respect to $t$. The conservation law for the congruence of the null rays can be represented as follows
\begin{equation}\label{64-label}
\begin{aligned}
& \nabla_\mu \, (\varrho_n \, K^\mu) = 0,\\
& \Rightarrow \quad \frac{1}{\varrho_n} \, \frac{d \varrho_n}{d \ell} = - \nabla_\mu K^\mu = - \frac{1}{\sqrt{- g}} \, \frac{\partial}{\partial x^\alpha} \, (\sqrt{- g} \, K^\alpha),
\end{aligned}
\end{equation}
where $\varrho_n$  is the density of null rays. According to Eq. \eqref{63-label}, we rewrite Eq. \eqref{64-label} as follows
\begin{equation}\label{65a-label}
\begin{aligned}
\frac{1}{\varrho_n}\,\frac{d\varrho_n}{d\ell}=&-\frac{1}{\sqrt{-g}}\,\frac{\partial}{\partial t}\,\big[\sqrt{-g}\,(1-\epsilon\,\delta_\ell^\prime)\big]-\frac{1}{\sqrt{-g}}\,\frac{\partial}{\partial r}\,\big[\sqrt{-g}\,\epsilon\,r_0\,\delta_r^\prime\big]\\
&-\frac{\Omega_\pm}{\sqrt{-g}}\,\frac{\partial}{\partial\phi}\,\big[\sqrt{-g}\,(1-\epsilon\,\delta_\ell^\prime)\big].
\end{aligned}
\end{equation} 
where $ g $ is the determinant of metric \eqref{50-label} and according to the metric, $ \sqrt{-g} $ is obtained as follows
\begin{equation}\label{65ba-label}
\sqrt{-g}=r^2+n^2+\nu\,r^2\,\ln(1+\frac{m^2}{r\,(r-2\,m)}).
\end{equation}
Using Eqs. \eqref{59-label} and \eqref{61-label}, we write $ \sqrt{-g} $ as below
\begin{equation}\label{65b-label}
\sqrt{-g}=\frac{19\,n^2}{3}+9\,m^2\,(1+2\,\epsilon\,\delta_r+\nu\,\ln\frac{4}{3}).
\end{equation}
According to Eqs. \eqref{65a-label} and \eqref{65b-label}, we have
\begin{equation}\label{65c-label}
\frac{1}{\varrho_n} \, \frac{d \varrho_n}{d \ell} = - \epsilon\,\big[\frac{\partial(2\,\delta_r-\delta_\ell^\prime)}{\partial t}+\frac{\partial(r_0\,\delta_r^\prime)}{\partial r}+\Omega_\pm\,\frac{\partial(2\,\delta_r-\delta_\ell^\prime)}{\partial\phi}\big].
\end{equation}
According to Eq. \eqref{61-label} ($dr/dt=\epsilon\,r_0\,\delta_r^\prime$), we have 
\begin{equation}\label{65d-label}
\frac{1}{\varrho_n} \, \frac{d \varrho_n}{d t} =\frac{d\,(\epsilon\,r_0\,\delta_r^\prime)}{dt}\,\frac{dt}{dr}+\mathcal{O}(\epsilon)= - \frac{\delta_r^{\prime \prime}(t)}{\delta_r^\prime(t)}+\mathcal{O}(\epsilon).
\end{equation} 
To obtain $ \varrho_n $, we must obtain $ \delta_r(t) $. To calculate $\delta_r$, we use the following equation, which is the radial component of geodesic
\begin{equation}\label{66-label}
\frac{d^2 r}{d \ell^2} + \Gamma_{t t}^r  \,\big( \frac{d t}{d \ell} \big)^2 + \Gamma_{\phi \phi}^r \, \big( \frac{d \phi}{d \ell} \big)^2 + \Gamma_{\theta \theta}^r \, \big( \frac{d \theta}{d \ell} \big)^2 = 0.
\end{equation}
By putting $ \theta=\pi/2 $ and expanding the above equation up to the first order of $ \epsilon $, we have
\begin{equation}\label{67-label}
\begin{aligned}
9\,m^2\,(27\,m^2+8\,n^2)\,\delta_r^{\prime\prime}(t)+\big[n^2-9\,m^2\,\big(1-\nu\,\ln(\frac{4}{3})\big)\big]\,\delta_r^{\prime}(t)=0.
\end{aligned}
\end{equation} 
By solving the differential Eq. \eqref{67-label}, $ \delta_r(t)$ is obtained as follows
\begin{equation}\label{68-label}
\delta_r(t) = \sinh (\zeta t),
\end{equation}
where $ \zeta $ is
\begin{equation}\label{69-label}
\zeta = \frac{1}{3\,\sqrt{3}\,m}\big[1-\frac{11\,n^2}{54\,m^2}-\nu\,\ln2+\frac{\nu\,\ln3}{2}\big].
\end{equation}
Using Eq. \eqref{68-label} in Eq. \eqref{65d-label}, $ \varrho_n $ is obtained as follows
\begin{equation}\label{70-label}
\varrho_n(t) = \varrho_n(0) \, \frac{1}{\cosh(\zeta)} \simeq 2 \, \varrho_n(0) \, (e^{- \zeta \, t} - e^{- 3 \, \zeta \, t} + e^{- 5 \, \zeta \, t} - \dots).
\end{equation}
Therefore, the imaginary parts related to the QNMs can be written as below
\begin{equation}\label{71-label}
\Gamma = \big( \tilde{n} + \frac{1}{2} \big) \, \zeta.
\end{equation}
QNMs for the TNS metric are given by
\begin{equation}\label{72-label}
\begin{aligned}
Q_{TNS}=&\big(\Omega+i\,\Gamma\big)_{TNS}=j\,\big[\frac{1}{3\,\sqrt{3}\,m}\,\big(1\pm\frac{5}{18}\,(\frac{n}{m})^2\big)\big]+i\,\big(\tilde{n}+\frac{1}{2}\big)\,\Big\{\frac{1}{3\,\sqrt{3}\,m}\big[1-\frac{11\,n^2}{54\,m^2}\\
&-\nu\,\ln2+\frac{\nu\,\ln3}{2}\big]\Big\}.
\end{aligned}
\end{equation}
\subsection{QNMs OF THE CTNS METRIC}\label{63}
Now, we derive the QNMs of the CTNS metric with the same method introduced in the previous section. To derive the QNMs, we first expand the metric in Eq. \eqref{eq3.24} to the second order of $N$ and the first order of $\nu$.
\begin{equation}\label{73-label}
\begin{aligned}
ds^2=&\Big\{\frac{(\tilde{r}-2M)^2}{3\,\tilde{r}^2}\,\big[\nu\,\ln(\tilde{r}-2\,M)-3\big]+\frac{8\,N^2}{\tilde{r}^6}\big[(M-\tilde{r})^2\,\big(\tilde{r}^2-2\,M(\tilde{r}-M)\big)\big]\Big\}\,dt^2\\
&+\Big\{\frac{1}{(M-\tilde{r})^4}\,\big[\tilde{r}^4+\frac{8\,N^2\,(M-\tilde{r})^2\,\big(\tilde{r}^2+2M(M-\tilde{r})\big)}{(\tilde{r}-2M)^2}\big]-\frac{\nu\,\tilde{r}^4}{12\,(M-\tilde{r})^4}\\
&\times\big[6\,\ln\big(\tilde{r}-2M\big)-\ln\big(\frac{\tilde{r}^2\,(\tilde{r}-2\,M)^2+4\,M^2\,(\tilde{r}-M)^2}{\tilde{r}^2}\big)\big]\Big\}\,dr^2\\
&+\frac{1}{(\tilde{r}-M)^2}\,\Big\{\tilde{r}^4+4\,N^2\,(\tilde{r}-M)^2+\nu\,\tilde{r}^4\,\big[\ln\big(\frac{\tilde{r}^2\,(\tilde{r}-2\,M)^2+4\,M^2\,(\tilde{r}-M)^2}{\tilde{r}^2}\big)\\
&-\ln(\tilde{r}-2\,M)\big]\Big\}\,d\theta^2+\Big\{4\,N^2+\frac{\tilde{r}^4}{3\,(\tilde{r}-M)^2}\big[3-\nu\,\ln(\tilde{r}-2\,M)\big]\Big\}\,d\phi^2.
\end{aligned}
\end{equation}
We perform all the calculations of the previous part for this metric, with the difference that the radius of the light ring for the conformal Schwarzschild metric is as follows
\begin{equation}\label{74-label}
r_{CS}=(3+\sqrt{3})\,M.
\end{equation}
Finally, we obtain the QNMs of the CTNS metric as follows
\begin{equation}\label{74a-label}
\begin{aligned}
Q_{CTNS}=&\big(\Omega+i\,\Gamma\big)_{CTNS}=j\,\big[\frac{1}{108\,\sqrt{3}\,M^3}\,(18\,M^2\pm5\,N^2)\big]+i\,\big(n+\frac{1}{2}\big)\,\Big\{-\frac{11 N^2}{54 M^3}\\
&+ \frac{ 1 - 0.144 \nu}{M}\Big\}.
\end{aligned}
\end{equation}
According to Eqs. \eqref{72-label} and \eqref{74a-label} and the data in Table \ref{1tab}, for all values of $ (n=N) $, $ (m=M) $ and $\nu$, the ratio of $ \Gamma_{TNS} $ to $ \Gamma_{CTNS} $ is always constant and smaller than one. Therefore, the perturbations created in the metric $ \Gamma_{CTNS} $ disappear faster than the perturbations created in the metric $ \Gamma_{TNS} $.
\begin{table}
\centering
\caption{Numerical values of the imaginary part of the QNMs of TNS and CTNS metrics}
\label{1tab}
\begin{tabular}{|c|c|c|c|}
\hline
$ n,\,N $ & $ m,\,M $ & $\nu$ &$\frac{\Gamma_{TNS}}{\Gamma_{CTNS}}$ \\
\hline  
1 & 1 & -1 & 0.19\\
\hline 
\end{tabular}
\end{table}\\

\section{GRAVITATIONAL LENSING IN TNS METRICS}\label{sec:Lensing}
Gravitational lensing is a powerful method of understanding the universe and compact objects within it. So far, many studies have been done on gravitational lensing, for example, see Refs. \cite{schneider1992gravitational, wambsganss1998gravitational, narayan1999gravitational, schneider2006gravitational, gibbons2008applications}.

In this section, we are going to study the deflection angle of light in a gravitational lens in the background of metrics \eqref{h18_label} and then study the effect of parameters $ n $ and $\nu$ on the deflection angle. For this, first, we consider the optical metric \eqref{h18_label} (metric in $ \theta=\pi/2 $)
\begin{equation}\label{l49_label}
d t^2 = \tilde{g}_{rr} \, d r^2 + \tilde{g}_{\phi\phi} \, d\phi^2,
\end{equation}
According to the Eqs. \eqref{h18_label} and \eqref{l49_label}, $\tilde{g}_{rr}$ and $\tilde{g}_{\phi\phi}$ are defined as follows
\begin{equation}\label{l50_label}
\tilde{g}_{rr}=\frac{g_{r r}}{g_{tt}} =\frac{\Sigma^\nu}{\Delta^2},\qquad\tilde{g}_{\phi\phi}=\frac{g_{\phi\phi}}{g_{tt}} =\frac{r^2\,\chi}{\Delta^2},
\end{equation}
where $\Delta$, $\Sigma$ and $\chi$ are as below
\begin{equation}\label{l51_label}
\begin{aligned}
&\Delta=1-\frac{2\,(n^2+m\,r)}{n^2+r^2},\\
&\Sigma=\frac{(m-r)^2}{r\,(r-2\,m)-n^2},\\
&\chi=1-\frac{2\,m}{r}-\frac{n^2}{r^2}.
\end{aligned}
\end{equation}
In the phenomenon of gravitational lensing, deflection angle of light is obtained by the following relation
\begin{equation}\label{l53_label}
\delta = - \, \int_{0}^{\pi} \, \int_{r_0}^{\infty} \, \sqrt{\tilde{g}} \, \mathcal{K} \, d r \, d \phi,
\end{equation}
where $r_0$ is considered as the minimum distance from the source of gravity and $ \mathcal{K} $ is Gaussian curvature and is equal to
\begin{equation}\label{l51a_label}
\mathcal{K} = - \, \frac{1}{\sqrt{\tilde{g}}} \, \Big[ \partial_r \, \Big( \frac{1}{\sqrt{\tilde{g}_{rr}}} \, \partial_r \, \sqrt{\tilde{g}_{\phi\phi}} \Big) + \partial_\phi \, \Big( \frac{1}{\sqrt{\tilde{g}_{\phi\phi}}} \, \partial_\phi \, \sqrt{\tilde{g}_{rr}} \Big) \Big],
\end{equation}
According to the Eq. \eqref{l49_label}, $ \sqrt{\tilde{g}} $ is calculated as follows
\begin{equation}\label{52_label}
\sqrt{\tilde{g}} = \frac{r\,\chi^{\frac{1}{2}}\,\Sigma^{\frac{\nu}{2}}}{\Delta^2}.
\end{equation}
Using Eqs. \eqref{l50_label}, \eqref{l53_label} and \eqref{l51a_label}, we obtain the deflection angle as follows
\begin{equation}\label{l55_label}
\delta = - \, \int_{0}^{\pi} \, \int_{r_0}^{\infty} \, \partial_r \, \big[ \chi^\frac{1}{2}\,\Sigma^{-\frac{\nu}{2}}\,(1-r\,\frac{\Delta^\prime}{\Delta}+\frac{r}{2}\,\frac{\chi^\prime}{\chi}) \big] \, d r \, d \phi.
\end{equation}
Using Eqs. \eqref{l51_label} in Eq. \eqref{l55_label} and expanding the obtained result up to order $r^{-2}$, we have
\begin{equation}\label{l57_label}
\begin{aligned}
\delta=&-\int_{0}^{\pi} \, \int_{r_0}^{\infty} \,\partial_r\,\big[1-\frac{m^2\,(\nu+3)+n^2\,(\nu+7)+4\,m\,r}{2\,r^2}\big]\,dr\,d\phi.
\end{aligned}
\end{equation}
To calculate $ \delta $, we must first obtain $r_0$, for this we first write metric \eqref{h18_label} in the equatorial plane ($ \theta=\pi/2 $) as follows
\begin{equation}\label{l58_label}
ds^2 = - \Delta \, dt^2 + \frac{\Sigma^\nu}{\Delta} \, dr^2 + \frac{r^2 \, \chi}{\Delta} \, d \phi^2,
\end{equation}
Also, we write geodesics Lagrangian as below
\begin{equation}\label{l60_label}
\mathcal{L} = - \frac{1}{2} \, \Delta \, \dot{t}^2 + \frac{1}{2} \, \frac{\Sigma^\nu}{\Delta} \, \dot{r}^2 + \frac{1}{2} \, \frac{r^2 \, \chi}{\Delta} \, \dot{\phi}^2,
\end{equation}
where dot is the derivative with respect to the affine parameter ($\lambda$). In the following, we obtain the equations of motion related to Lagrangian \eqref{l60_label}
\begin{equation}\label{l61_label}
E =\Delta \, \frac{d t}{d \lambda} =  const,
\end{equation}
\begin{equation}\label{l62_label}
L = \frac{r^2 \, \chi}{\Delta} \, \frac{d \phi}{d \lambda} = const.
\end{equation}
By dividing Eq. \eqref{l61_label} by Eq. \eqref{l62_label}, we have
\begin{equation}\label{l63_label}
\frac{E}{L} =\frac{\Delta^2}{r^2\,\chi} \, \frac{d t}{d \phi} = \frac{1}{b},
\end{equation}
that $ b $ is a constant. According to metric \eqref{l58_label}, we have for light 
\begin{equation}\label{l64_label}
ds^2=0,\quad\Rightarrow\quad\frac{\Sigma^\nu}{\Delta} \, \big( \frac{d r}{d \lambda} \big)^2 = \Delta \, \big( \frac{d t}{d \lambda} \big)^2 - \frac{r^2 \, \chi}{\Delta} \, \big( \frac{d \phi}{d \lambda} \big)^2,
\end{equation}
where by using $ d\phi $ instead of $ d\lambda $, we get the following equation
\begin{equation}\label{l65_label}
\frac{\Sigma^\nu}{\Delta} \, \big( \frac{d r}{d \phi} \big)^2 = \frac{r^4\,\chi^2}{b^2\,\Delta} - \frac{r^2 \, \chi}{\Delta}.
\end{equation}
Now we use the change of variable $ u=1/r $ and rewrite the Eq. \eqref{l65_label} as below
\begin{equation}\label{l67_label}
\big( \frac{d u}{d \phi} \big)^2 = \chi\,\Sigma^{-\nu} \, \big( \frac{\chi}{b^2} - u^2\big).
\end{equation}
We derive Eq. \eqref{l67_label} with respect to $u$, then we expand the obtained relation to the order of $u^2$, which gives the result
\begin{equation}\label{l68_label}
\frac{d^2 u}{d \phi^2} + u = 3 \, m \, u^2+\frac{u\,\big[2\,n^2-\nu\,(m^2+n^2)\,(1+3\,m\,u)\big]}{b^2}.
\end{equation}
In the following, we first obtain the solution of the following differential equation
\begin{equation}\label{l69_label}
\frac{d^2 u}{d \phi^2} + u = 0, \quad \Rightarrow\quad u = \frac{\sin \phi}{b}.
\end{equation}
Also, Eq. \eqref{l68_label}, up to the order of $ u^2 $ is written as follows
\begin{equation}\label{l71_label}
\frac{d^2 u}{d \phi^2} + u - 3 \, m \, u^2 = 0.
\end{equation}
Finally, considering $u$ as follows
\begin{equation}\label{l72a_label}
u = \frac{\sin \phi}{b}+ F(\phi),
\end{equation}
and solving the differential equation \eqref{l71_label}, we have
\begin{equation}\label{l72_label}
u = \frac{\sin \phi}{b}+ \frac{m}{b^2} \, (1 + \cos^2 \phi).
\end{equation}
Finally, by changing the variable $ u=1/r $ in Eq. \eqref{l57_label} and using Eq. \eqref{l72_label} we have
\begin{equation}\label{l73_label}
\delta_{TNS}=\frac{4\,m}{b}+\frac{\pi}{4\,b^2}\,\big[m^2\,(\nu+15)+n^2\,(\nu+7)\big].
\end{equation}
Now, if $n = 0$ and $\nu = 0$ in Eq. \eqref{l73_label}, we have
\begin{equation}\label{l74_label}
\delta=\frac{4\,m}{b}+\frac{15\,\pi\,m^2}{4\,b^2}.
\end{equation}
Eq. \eqref{l74_label} shows the  in the deflection angle in Schwarzschild metric.

Now we want to investigate the effect of NUT parameter and scalar field on deflection angle of light. To study this issue, we obtain the ratio of the deflection angle in TNS metrics to Taub-NUT metrics ($ \nu=0 $) and metrics in the presence of the scalar field ($ n=0 $), which are respectively as follows
\begin{equation}\label{l75_label}
\frac{\delta_{TNS}}{\delta_{TN}}=1+\frac{\pi\,\nu\,(m^2+n^2)}{16\,m\,b+15\,\pi\,m^2+7\,\pi\,n^2},
\end{equation}
\begin{equation}\label{l76_label}
\frac{\delta_{TNS}}{\delta_S}=1+\frac{\pi\,n^2\,(\nu+7)}{m\,\big[16\,b+\pi\,m\,(\nu+15)\big]}.
\end{equation}
In Eq. \eqref{l75_label}, despite the fact that $\nu$ is always negative, the deflection angle of light in TNS metrics is always lower than the deflection angle of light in Taub-NUT metrics. So we conclude that the presence of the scalar field in Taub-NUT metrics causes the deflection angle of light to decrease.

In Eq. \eqref{l76_label}, considering that $b$ is a large value ($b>m$), we have the following conditions
\begin{equation}\label{5b_label}
\frac{\delta_{TNS}}{\delta_S}\begin{cases}
>1,\qquad for\; \nu>-7,\\=0,\qquad for\; \nu=-7,\\<1,\qquad for\; \nu<-7.\\
\end{cases}
\end{equation}
Therefore, according to Eq. \eqref{5b_label}, the deflection angle of light in TNS metrics to the deflection angle of light in metrics in the presence of a scalar field can be smaller, equal, or greater than one depending on the value of $\nu$.

\section*{CONCLUSION}\label{sec:con}
Using three different methods i.e. the Ernst potential and equations and two types of Ehlers transformations, we obtained a new class of exact solutions for Einstein's field equations. Using the topological charge method, the light ring in the Taub-NUT-scalar (TNS) geometry were investigated in such a way that the presence of the topological charge indicates the existence of the light ring in this new geometry, and the topological charge location, with a good approximation, was the extreme point of the effective potential. Furthermore, by using the appropriate conformal transformation, we obtained an exact class of solutions of the conformal-Taub-NUT-scalar (CTNS) theory. We also compared the effective potentials of TNS and CTNS spece-times. By comparing the QNMs of the TNS and CTNS metrics, we concluded that small perturbations in the TNS space-time are more stable than the small perturbations in the CTNS space-time. Gravitational lensing is also studied for different class of metrics. There are other different aspect of the class of metrics that can be studied in the future. 
\section*{APPENDIX A}
$ X(r,\,\theta) $ in Eq. \eqref{k1} is determined by the following equation. It should also be noted that $ X(r,\,\theta) $ is a regular function of $ r $ and $ \theta $ and does not have a singularity at any point.
\begin{equation}\label{app1}
\begin{aligned}
X(r,\,\theta)=&\frac{1}{\big[(n^2+m^2)\,\cos^2\theta-(m-r)^2\big]^{2\,\nu+1}}\Big\{3\,(m^2+n^2)\,\cos^2\theta\,\Big\{\nu^2\,(n^2+r^2)^2\\
&\times(m^2+n^2)\,\Big[n^4\,\big(m^2-\frac{20}{3}\,r\,m+\frac{22}{3}\,r^2\big)+n^2\,\big(\frac{40}{3}\,m\,r^3-3\,r^4-6\,m^2\,r^2\big)\\
&+\frac{11}{3}\,m\,r^4\,\big(m-\frac{4}{11}\,r\big)-\frac{n^6}{3}\Big]-24\,\nu\,(n^2+r^2)\,\big[\frac{n^2}{2}+r\,(m-\frac{r}{2})\big]\Big[n^6\,\big(\frac{4}{9}\,m^2\\
&-3\,r\,m+\frac{19}{9}\,r^2\big)+r\,n^4\,(m-r)\,\big(m^2-\frac{62}{9}\,r\,m+\frac{8}{3}\,r^2\big)+n^2\,\big(\frac{r^6}{3}-\frac{31}{9}\,m\,r^5\\
&+\frac{22}{3}\,m^2\,r^4-\frac{34}{9}\,m^3\,r^3\big)+\frac{5}{9}\,m^2\,r^5\,\big(m-\frac{3}{5}\,r\big)-\frac{2}{9}\,n^8\Big]-16\,\big[\frac{n^2}{2}+r\,\big(m\\
&-\frac{r}{2}\big)\big]^2\big[n^4+n^3\,(m-3\,r)+3\,n^2\,r\,(m-r)+n\,r^2\,(r-3\,m)-m\,r^3\big]\,\big[n^3\,(m\\
&-3\,r)-3\,n^2\,r\,(m-r)+n\,r^2\,(r-3\,m)+m\,r^3-n^4\big]\Big\}-3\,\nu^2\,(n^2+r^2)^2\\
&\times(m^2+n^2)^2\,\Big[r^6-\frac{10}{3}\,m\,r^5+\frac{11}{3}\,m^2\,r^4-\frac{4}{3}\,n^6+n^4\,\big(m^2-\frac{26}{3}\,r\,m\\
&+\frac{19}{3}\,r^2\big)+2\,n^2\,r^2\,\big(\frac{14}{3}\,m\,r-r^2-3\,m^2\big)\Big]+72\,\nu\,(n^2+r^2)\,(m^2+n^2)\\
&\times(m-r)\,\big[\frac{n^2}{2}+r\,\big(m-\frac{r}{2}\big)\big]\,\Big[\frac{n^6}{9}\,\big(4\,m-13\,r\big)+n^4\,r\,\big(m^2-\frac{50}{9}\,r\,m+\frac{26}{9}\,r^2\big)\\
&-n^2\,r^3\,\big(r^2-\frac{40}{9}\,m\,r+\frac{34}{9}\,m^2\big)+\frac{5}{9}\,m\,r^5\,\big(m-\frac{2}{5}\,r\big)\Big]-48\,(m-r)^2\,\big[\frac{n^2}{2}\\
&+r\,\big(m-\frac{r}{2}\big)\big]^2\,\big[n^4+n^3\,(m-3\,r)+3\,n^2\,r\,(m-r)+n\,r^2\,(r-3\,m)\\
&-m\,r^3\big]\,\big[n^4-n^3\,(m-3\,r)+3\,n^2\,r\,(m-r)-n\,r^2\,(r-3\,m)-m\,r^3\big]\Big\}.
\end{aligned}
\end{equation}


\begin{thebibliography}{00}
	
	\bibitem{Mirza2023}
	A. Azizallahi, B. Mirza, A. Hajibarat, H. Anjomshoa, Three parameter metrics in the presence of a scalar field in four and higher dimensions, 	\hyperref{https://doi.org/10.1016/j.nuclphysb.2023.116414}{}{}{\color{blue}{Nuclear Physics B. 998, 116414 (2023)}}, \hyperref{https://arxiv.org/abs/2307.09328}{}{}{\color{blue}{arXiv:2307.09328 [gr-qc]}}.
	
	
	\bibitem{Mirza2023-2}
	B. Mirza, P. K. Kangazi, F. Sadeghi, A class of rotating metrics in the presence of a scalar field, \hyperref{https://doi.org/10.1140/epjc/s10052-023-12255-7}{}{}{\color{blue}{Eur. Phys. J. C. 83, 1161 (2023)}}, \hyperref{https://arxiv.org/abs/2307.13588}{}{}{\color{blue}{arXiv:2307.13588 [gr-qc]}}.
	
	
	\bibitem{Penrose1969-2}
	R. Penrose, Gravitational collapse: The role of general relativity, \hyperref{https://ui.adsabs.harvard.edu/abs/1969NCimR...1..252P/abstract}{}{}{\color{blue}{Nuovo Cimento Rivista Serie. 1, 252 (1969)}}.
	
	
	
	\bibitem{Hawking-1}
	S. Hawking, W. Israel, and D. Liebscher, General relativity: an Einstein centenary survey, \hyperref{https://ui.adsabs.harvard.edu/abs/2010grae.book.....H/abstract}{}{}{\color{blue}{Astron. Nachrichten. 301, 331 (1980)}}.
	
	
	\bibitem{Darmois1927}
	G. Darmois, Les équations de la gravitation einsteinienne, \hyperref{$ http://www.numdam.org/item/MSM_1927__25__1_0.pdf $}{}{}{\color{blue}{Mémorial des sciences mathématiques, no. 25 (1927), 58 p}}.
	
	
	\bibitem{erez1959gravitational}
	Erez, G and Rosen, N, The gravitational field of a particle possessing a multipole moment, \hyperref{https://www.osti.gov/biblio/4201189}{}{}{\color{blue}{Israel Inst. of Tech., Haifa. (1959)}}.
	
	
	
	\bibitem{Zipoy1966}
	D. M. Zipoy, Topology of some spheroidal metrics, \hyperref{https://ui.adsabs.harvard.edu/link-gateway/1966JMP.....7.1137Z/doi:10.1063/1.1705005}{}{}{\color{blue}{J. Math. Phys. 7, 1137 (1966)}}.
	
	
	
	\bibitem{Voorhees1970}
	B. Voorhees, Static axially symmetric gravitational fields, \hyperref{https://link.aps.org/doi/10.1103/PhysRevD.2.2119}{}{}{\color{blue}{Phys. Rev. D. 2, 2119 (1970)}}.
	
	
	\bibitem{Destounis2023}
	K. Destounis, G. Huez, and K.D. Kokkotas, Geodesics and gravitational waves in chaotic extreme-mass-ratio inspirals: the curious case of Zipoy-Voorhees black-hole mimickers, \hyperref{https://doi.org/10.1007/s10714-023-03119-2}{}{}{\color{blue}{Gen. Relativ. Gravit. 55, 71 (2023)}}, \hyperref{https://arxiv.org/abs/2301.11483}{}{}{\color{blue}{arXiv:2301.11483 [gr-qc]}}.
	
	
	\bibitem{Clavijo2023}
	F. D. Lora-Clavijo, G. D. Prada-Méndez, L. M. Becerra, and E. A. Becerra-Vergara, The q-metric naked singularity: a viable explanation for the nature of the central object in the Milky Way, \hyperref{https://dx.doi.org/10.1088/1361-6382/ad0b9e}{}{}{\color{blue}{Class. Quantum Grav. 40, 245012 (2023)}}, \hyperref{https://arxiv.org/abs/2311.06653}{}{}{\color{blue}{arXiv:2311.06653 [gr-qc]}}.
	
	
	
	\bibitem{herrera2000geodesics}
	Herrera, L and Paiva, Filipe M and Santos, NO, Geodesics in the $\gamma$-Spacetime, \hyperref{https://www.worldscientific.com/doi/abs/10.1142/S021827180000061X}{}{}{\color{blue}{International Journal of Modern Physics D. 9, 649--659 (2000)}}, \hyperref{https://arxiv.org/abs/gr-qc/9812023}{}{}{\color{blue}{arXiv:gr-qc/9812023}}.
	
	
	
	\bibitem{richterek2002einstein}
	Richterek, Luk{\'a}{\v{s}} and Novotn{\`y}, Jan and Horsk{\`y}, Jan, Einstein-Maxwell fields generated from the $\gamma$-metric and their limits, \hyperref{https://link.springer.com/article/10.1023/A:1020581415399}{}{}{\color{blue}{Czechoslovak journal of physics. 52, 1021--1040 (2002)}}, \hyperref{https://arxiv.org/abs/gr-qc/0209094}{}{}{\color{blue}{arXiv:gr-qc/0209094}}.
	
	
	\bibitem{chakrabarty2018unattainable}
	Chakrabarty, Hrishikesh and Benavides-Gallego, Carlos A and Bambi, Cosimo and Modesto, Leonardo, Unattainable extended spacetime regions in conformal gravity, \hyperref{https://link.springer.com/article/10.1007/JHEP03(2018)013}{}{}{\color{blue}{Journal of High Energy Physics. 2018, 1--12 (2018)}}, \hyperref{https://arxiv.org/abs/1711.07198}{}{}{\color{blue}{arXiv:1711.07198 [gr-qc]}}.
	
	
	
	\bibitem{abdikamalov2019black}
	Abdikamalov, Askar B and Abdujabbarov, Ahmadjon A and Ayzenberg, Dimitry and Malafarina, Daniele and Bambi, Cosimo and Ahmedov, Bobomurat, Black hole mimicker hiding in the shadow: Optical properties of the $\gamma$-metric, \hyperref{https://journals.aps.org/prd/abstract/10.1103/PhysRevD.100.024014}{}{}{\color{blue}{Physical Review D. 100, 024014 (2019)}}, \hyperref{https://arxiv.org/abs/1904.06207}{}{}{\color{blue}{arXiv:1904.06207 [gr-qc]}}.
	
	
	\bibitem{toshmatov2019harmonic}
	Toshmatov, Bobir and Malafarina, Daniele and Dadhich, Naresh, Harmonic oscillations of neutral particles in the $\gamma$-metric, \hyperref{https://journals.aps.org/prd/abstract/10.1103/PhysRevD.100.044001}{}{}{\color{blue}{Physical Review D. 100, 044001 (2019)}}, \hyperref{https://arxiv.org/abs/1905.01088}{}{}{\color{blue}{arXiv:1905.01088 [gr-qc]}}.
	
	
	\bibitem{benavides2019charged}
	Benavides-Gallego, Carlos A and Abdujabbarov, Ahmadjon and Malafarina, Daniele and Ahmedov, Bobomurat and Bambi, Cosimo, Charged particle motion and electromagnetic field in $\gamma$-spacetime, \hyperref{https://journals.aps.org/prd/abstract/10.1103/PhysRevD.99.044012}{}{}{\color{blue}{Physical Review D. 99, 044012 (2019)}}, \hyperref{https://arxiv.org/abs/1812.04846}{}{}{\color{blue}{arXiv:1812.04846 [gr-qc]}}.
	
	
	\bibitem{allahyari2019quasinormal}
	Allahyari, Alireza and Firouzjahi, Hassan and Mashhoon, Bahram, Quasinormal modes of a black hole with quadrupole moment, \hyperref{https://journals.aps.org/prd/abstract/10.1103/PhysRevD.99.044005}{}{}{\color{blue}{Physical Review D. 99, 044005 (2019)}}, \hyperref{https://arxiv.org/abs/1812.03376}{}{}{\color{blue}{arXiv:1812.03376 [gr-qc]}}.
	
	
	
	\bibitem{allahyari2020quasinormal}
	Allahyari, Alireza and Firouzjahi, Hassan and Mashhoon, Bahram, Quasinormal modes of generalized black holes: $\delta$-Kerr spacetime, \hyperref{https://iopscience.iop.org/article/10.1088/1361-6382/ab6860/meta}{}{}{\color{blue}{Classical and Quantum Gravity. 37, 055006 (2020)}}, \hyperref{https://arxiv.org/abs/1908.10813}{}{}{\color{blue}{arXiv:1908.10813 [gr-qc]}}.
	
	
	\bibitem{chakrabarty2022effects}
	Chakrabarty, Hrishikesh and Borah, Debasish and Abdujabbarov, Ahmadjon and Malafarina, Daniele and Ahmedov, Bobomurat, Effects of gravitational lensing on neutrino oscillation in $\gamma$-spacetime, \hyperref{https://link.springer.com/article/10.1140/epjc/s10052-021-09982-0}{}{}{\color{blue}{The European Physical Journal C. 82, 24 (2022)}}, \hyperref{https://arxiv.org/abs/2109.02395}{}{}{\color{blue}{arXiv:2109.02395 [gr-qc]}}.
	
	
	
	\bibitem{hajibarat2022gamma}
	Hajibarat, Arash and Mirza, Behrouz and Azizallahi, Alireza, $\gamma$-metrics in higher dimensions, \hyperref{https://www.sciencedirect.com/science/article/pii/S0550321322000906}{}{}{\color{blue}{Nuclear Physics B. 978, 115739 (2022)}}, \hyperref{https://arxiv.org/abs/2110.06667}{}{}{\color{blue}{	arXiv:2110.06667 [gr-qc]}}.
	
	
	\bibitem{li2022constraining}
	Li, Song and Mirzaev, Temurbek and Abdujabbarov, Ahmadjon A and Malafarina, Daniele and Ahmedov, Bobomurat and Han, Wen-Biao, Constraining the deformation of a rotating black hole mimicker from its shadow, \hyperref{https://journals.aps.org/prd/abstract/10.1103/PhysRevD.106.084041}{}{}{\color{blue}{Physical Review D. 106, 084041 (2022)}}, \hyperref{https://arxiv.org/abs/2207.10933}{}{}{\color{blue}{arXiv:2207.10933 [gr-qc]}}.
	
	
	
	\bibitem{chakrabarty2023constraining}
	Chakrabarty, Hrishikesh and Tang, Yong, Constraining deviations from spherical symmetry using $\gamma$-metric, \hyperref{https://journals.aps.org/prd/abstract/10.1103/PhysRevD.107.084020}{}{}{\color{blue}{Physical Review D. 107, 084020 (2023)}}, \hyperref{https://arxiv.org/abs/2204.06807}{}{}{\color{blue}{arXiv:2204.06807 [gr-qc]}}.

	
	
	\bibitem{harada2002physical}
	Harada, Tomohiro and Iguchi, Hideo and Nakao, Ken-ichi, Physical processes in naked singularity formation, \hyperref{https://academic.oup.com/ptp/article/107/3/449/1809530}{}{}{\color{blue}{Progress of Theoretical Physics. 107, 449--524 (2002)}}, \hyperref{https://arxiv.org/abs/gr-qc/0204008v1}{}{}{\color{blue}{arXiv:gr-qc/0204008}}.
	
	
	
	\bibitem{Fisher1948}
	I. Fisher, Scalar mesostatic field with regard for gravitational effects, \hyperref{https://doi.org/10.48550/arXiv.gr-qc/9911008}{}{}{\color{blue}{Zh. Eksp. Teor. Fiz. 18, 636-640 (1948)}}.
	
	
	\bibitem{Janis1968}
	A. I. Janis, E. T. Newman, and J. Winicour, Reality of the Schwarzschild singularity, \hyperref{https://link.aps.org/doi/10.1103/PhysRevLett.20.878}{}{}{\color{blue}{Phys. Rev. Lett. 20, 878 (1968)}}.
	
	
	
	\bibitem{Wyman1981}
	M. Wyman, Static spherically symmetric scalar fields in general relativity, \hyperref{https://link.aps.org/doi/10.1103/PhysRevD.24.839}{}{}{\color{blue}{Phys. Rev. D. 24, 839 (1981)}}.
	
	
	
	
	\bibitem{Virbhadra1998}
	K. Virbhadra, D. Narasimha, and S. Chitre, Role of the scalar field in gravitational lensing, \hyperref{https://doi.org/10.48550/arXiv.astro-ph/9801174}{}{}{\color{blue}{Astron. Astrophys. 337, 1-8 (1998)}}, \hyperref{https://arxiv.org/abs/astro-ph/9801174}{}{}{\color{blue}{arXiv:astro-ph/9801174}}.
	
	
	\bibitem{Dey2008}
	T. K. Dey and S. Sen, Gravitational lensing by wormholes, \hyperref{https://doi.org/10.1142/S0217732308025498}{}{}{\color{blue}{Mod. Phys. Lett. 23, 953 (2008)}}, \hyperref{https://arxiv.org/abs/gr-qc/0602062v4}{}{}{\color{blue}{arXiv:gr-qc/0602062}}.
	
	
	
	\bibitem{Chowdhury2012}
	A. N. Chowdhury, M. Patil, D. Malafarina, and P. S. Joshi, Circular geodesics and accretion disks in the Janis-Newman-Winicour and gamma metric spacetimes, \hyperref{https://link.aps.org/doi/10.1103/PhysRevD.85.104031}{}{}{\color{blue}{Phys. Rev. D. 85, 104031 (2012)}}, \hyperref{https://arxiv.org/abs/1112.2522}{}{}{\color{blue}{arXiv:1112.2522 [gr-qc]}}.
	
	
	\bibitem{Turimov2018}
	B. Turimov, B. Ahmedov, M. Kološ, and Z. Stuchlík, Axially symmetric and static solutions of Einstein equations with self-gravitating scalar field, \hyperref{https://link.aps.org/doi/10.1103/PhysRevD.98.084039}{}{}{\color{blue}{Phys. Rev. D. 98, 084039 (2018)}}, \hyperref{https://arxiv.org/abs/1810.01460}{}{}{\color{blue}{arXiv:1810.01460 [gr-qc]}}.
	
	
	
	
	\bibitem{taub1951empty}
	Taub, Abraham H, Empty space-times admitting a three parameter group of motions, \hyperref{https://www.jstor.org/stable/1969567}{}{}{\color{blue}{Annals of Mathematics. 53, 472--490 (1951)}}.
	

	
	\bibitem{newman1963empty}
	Newman, Ezra and Tamburino, L and Unti, Theodore, Empty-space generalization of the Schwarzschild metric, \hyperref{https://pubs.aip.org/aip/jmp/article-abstract/4/7/915/230250/Empty-Space-Generalization-of-the-Schwarzschild}{}{}{\color{blue}{Journal of Mathematical Physics. 4, 915--923 (1963)}}.
	
	
	
	\bibitem{Kramer1983Exact}
	D. Kramer and H. Stephani, Exact solutions of Einstein’s field equations, \hyperref{https://ui.adsabs.harvard.edu/abs/1983grg..conf...75K/abstract}{}{}{\color{blue}{General Relativity
			and Gravitation. 1980, 75 (1983)}}.
	
	
	
	\bibitem{misner1963flatter}
	Misner, Charles W, The flatter regions of Newman, Unti, and Tamburino's generalized Schwarzschild space, \hyperref{https://pubs.aip.org/aip/jmp/article-abstract/4/7/924/230241/The-Flatter-Regions-of-Newman-Unti-and-Tamburino-s}{}{}{\color{blue}{Journal of Mathematical Physics. 4, 924--937 (1963)}}.
	
	
	\bibitem{misner1967contribution}
	Misner, CW, Contribution to lectures in applied mathematics, \hyperref{https://bookstore.ams.org/lam}{}{}{\color{blue}{American Mathematical Society. 8, 160 (1967)}}.
	
	
	
	\bibitem{bonnor1969new}
	Bonnor, William B, A new interpretation of the NUT metric in general relativity, \hyperref{https://www.cambridge.org/core/journals/mathematical-proceedings-of-the-cambridge-philosophical-society/article/abs/new-interpretation-of-the-nut-metric-in-general-relativity/3C7F5F3E0DC2F1B355F2B0D195EEE0D1}{}{}{\color{blue}{Cambridge University Press. 66, 145--151 (1969)}}.
	
	
	\bibitem{israel1977line}
	Israel, Werner, Line sources in general relativity, \hyperref{https://journals.aps.org/prd/abstract/10.1103/PhysRevD.15.935}{}{}{\color{blue}{Physical Review D. 15, 935 (1977)}}.
	
	
	\bibitem{manko2005physical}
	Manko, VS and Ruiz, E, Physical interpretation of the NUT family of solutions, \hyperref{https://iopscience.iop.org/article/10.1088/0264-9381/22/17/014/meta}{}{}{\color{blue}{Classical and Quantum Gravity. 22, 3555 (2005)}}, \hyperref{https://arxiv.org/abs/gr-qc/0505001v1}{}{}{\color{blue}{arXiv:gr-qc/0505001}}.
	
	
	
	\bibitem{gonzalez2018new}
	Gonz{\'a}lez, Hern{\'a}n A and Grumiller, Daniel and Merbis, Wout and Wutte, Raphaela, New entropy formula for Kerr black holes, \hyperref{https://ui.adsabs.harvard.edu/abs/2018EPJWC.16801009G/abstract}{}{}{\color{blue}{EPJ Web of Conferences. 168, 01009 (2018)}}, \hyperref{https://arxiv.org/abs/1709.09667}{}{}{\color{blue}{arXiv:1709.09667 [hep-th]}}.
	
	
	
	\bibitem{hennigar2019thermodynamics}
	Hennigar, Robie A and Kubiz{\v{n}}{\'a}k, David and Mann, Robert B, Thermodynamics of Lorentzian Taub-NUT spacetimes, \hyperref{https://journals.aps.org/prd/abstract/10.1103/PhysRevD.100.064055}{}{}{\color{blue}{Physical Review D. 100, 064055 (2019)}}, \hyperref{https://arxiv.org/abs/1903.08668}{}{}{\color{blue}{arXiv:1903.08668 [hep-th]}}.
	
	
	
	\bibitem{bordo2019first}
	Bordo, Alvaro Ballon and Gray, Finnian and Hennigar, Robie A and Kubiz{\v{n}}{\'a}k, David, The first law for rotating NUTs, \hyperref{https://www.sciencedirect.com/science/article/pii/S037026931930694X}{}{}{\color{blue}{Physics Letters B. 798, 134972 (2019)}}, \hyperref{https://arxiv.org/abs/1905.06350}{}{}{\color{blue}{arXiv:1905.06350 [hep-th]}}.
	
	
	
	\bibitem{bordo2019thermodynamics}
	Bordo, Alvaro Ballon and Gray, Finnian and Kubiz{\v{n}}{\'a}k, David, Thermodynamics and phase transitions of NUTty dyons, \hyperref{https://link.springer.com/article/10.1007/JHEP07(2019)119}{}{}{\color{blue}{Journal of High Energy Physics. 2019, 1--21 (2019)}}, \hyperref{https://arxiv.org/abs/1904.00030}{}{}{\color{blue}{arXiv:1904.00030 [hep-th]}}.
	
	
	
	
	\bibitem{bordo2019misner}
	Bordo, Alvaro Ballon and Gray, Finnian and Hennigar, Robie A and Kubiz{\v{n}}{\'a}k, David, Misner gravitational charges and variable string strengths, \hyperref{https://iopscience.iop.org/article/10.1088/1361-6382/ab3d4d/meta}{}{}{\color{blue}{Classical and Quantum Gravity. 36, 194001 (2019)}}, \hyperref{https://arxiv.org/abs/1905.03785}{}{}{\color{blue}{arXiv:1905.03785 [hep-th]}}.
	
	
	
	
	\bibitem{zhang2021nut}
	Zhang, Ming and Jiang, Jie, NUT charges and black hole shadows, \hyperref{https://www.sciencedirect.com/science/article/pii/S0370269321001532}{}{}{\color{blue}{Physics Letters B. 816, 136213 (2021)}}, \hyperref{https://arxiv.org/abs/2103.11416}{}{}{\color{blue}{arXiv:2103.11416 [gr-qc]}}.
	
	
	
	\bibitem{arratia2021hairy}
	Arratia, Esteban and Corral, Crist{\'o}bal and Figueroa, Jos{\'e} and Sanhueza, Leonardo, Hairy Taub-NUT/bolt-AdS solutions in Horndeski theory, \hyperref{https://journals.aps.org/prd/abstract/10.1103/PhysRevD.103.064068}{}{}{\color{blue}{Physical Review D. 103, 064068 (2021)}}, \hyperref{https://arxiv.org/abs/2010.02460}{}{}{\color{blue}{arXiv:2010.02460 [hep-th]}}.
	
	
	\bibitem{barrientos2022gravitational}
	Barrientos, Jos{\'e} and Cisterna, Adolfo and Corral, Crist{\'o}bal and Oyarzo, Marcelo, Gravitational instantons with conformally coupled scalar fields, \hyperref{https://link.springer.com/article/10.1007/JHEP05(2022)110}{}{}{\color{blue}{Journal of High Energy Physics. 2022, 1--28 (2022)}}, \hyperref{https://arxiv.org/abs/2202.13854}{}{}{\color{blue}{arXiv:2202.13854 [hep-th]}}.
	
	
	
	
	\bibitem{durka2022first}
	Durka, Remigiusz, The first law of black hole thermodynamics for Taub--NUT spacetime, \hyperref{https://www.worldscientific.com/doi/abs/10.1142/S0218271822500213}{}{}{\color{blue}{International Journal of Modern Physics D. 31, 2250021 (2022)}}, \hyperref{https://arxiv.org/abs/1908.04238}{}{}{\color{blue}{arXiv:1908.04238 [gr-qc]}}.
	
	
	
	\bibitem{cano2022quasinormal}
	Cano, Pablo A and Pere{\~n}iguez, David, Quasinormal modes of NUT-charged black branes in the AdS/CFT correspondence, \hyperref{https://iopscience.iop.org/article/10.1088/1361-6382/ac7d8d/meta}{}{}{\color{blue}{Classical and Quantum Gravity. 39, 165003 (2022)}}, \hyperref{https://arxiv.org/abs/2101.10652}{}{}{\color{blue}{arXiv:2101.10652 [hep-th]}}.
	
	
	
	\bibitem{Ernst1968-1}
	F. J. Ernst, New formulation of the axially symmetric gravitational field problem, \hyperref{https://journals.aps.org/pr/abstract/10.1103/PhysRev.167.1175}{}{}{\color{blue}{Phys. Rev. 167, 1175 (1968)}}.
	
	\bibitem{Ernst1968-2}
	F. J. Ernst, New formulation of the axially symmetric gravitational field problem. II, \hyperref{https://journals.aps.org/pr/abstract/10.1103/PhysRev.172.1850.3}{}{}{\color{blue}{Phys. Rev. 168, 1415 (1968)}}.
	
	
	\bibitem{reina1976axisymmetric}
	Reina, C and Treves, A, Axisymmetric gravitational fields, \hyperref{https://link.springer.com/article/10.1007/BF00778761}{}{}{\color{blue}{General Relativity and Gravitation. 7, 817--838 (1976)}}.
	
	
	
	\bibitem{kinnersley1977symmetries1}
	Kinnersley, William, Symmetries of the stationary {E}instein--{M}axwell field equations. {I}, \hyperref{https://aip.scitation.org/doi/abs/10.1063/1.523458}{}{}{\color{blue}{J. Math. Phys. 18, 1529--1537 (1977)}}.
	
	\bibitem{kinnersley1977symmetries}
	Kinnersley, William and Chitre, DM, Symmetries of the stationary {E}instein--{M}axwell field equations. {II}, \hyperref{https://aip.scitation.org/doi/10.1063/1.523459}{}{}{\color{blue}{J. Math. Phys. 18, 1538--1542 (1977)}}.
	
	
	\bibitem{kinnersley1978symmetries1}
	Kinnersley, William and Chitre, DM, Symmetries of the stationary {E}instein--{M}axwell field equations. {III}, \hyperref{https://aip.scitation.org/doi/abs/10.1063/1.523912}{}{}{\color{blue}{J. Math. Phys. 19, 1926--1931 (1978)}}.
	
	
	\bibitem{kinnersley1978symmetries}
	Kinnersley, William and Chitre, DM, Symmetries of the stationary {E}instein--{M}axwell equations. {IV}. Transformations which preserve asymptotic flatness, \hyperref{https://aip.scitation.org/doi/abs/10.1063/1.523580}{}{}{\color{blue}{J. Math. Phys. 19, 2037--2042 (1978)}}.
	
	
	\bibitem{hoenselaers1979symmetries1}
	Hoenselaers, C, Symmetries of the stationary {E}instein--{M}axwell field equations. {V}, \hyperref{https://aip.scitation.org/doi/abs/10.1063/1.524057}{}{}{\color{blue}{J. Math. Phys. 20, 2526--2529 (1979)}}.
	
	
	\bibitem{hoenselaers1979symmetries}
	Hoenselaers, C and Kinnersley, William and Xanthopoulos, Basilis C, Symmetries of the stationary {E}instein--{M}axwell equations. {VI}. Transformations which generate asymptotically flat spacetimes with arbitrary multipole moments, \hyperref{https://aip.scitation.org/doi/abs/10.1063/1.524058}{}{}{\color{blue}{J. Math. Phys. 20, 2530--2536 (1979)}}.
	
	
	\bibitem{cosgrove1980relationships}
	Cosgrove, Christopher M, Relationships between the group-theoretic and soliton-theoretic techniques for generating stationary axisymmetric gravitational solutions, \hyperref{https://pubs.aip.org/aip/jmp/article-abstract/21/9/2417/225319/Relationships-between-the-group-theoretic-and}{}{}{\color{blue}{J. Math. Phys. 21, 2417--2447 (1980)}}.
	
	
	\bibitem{wu2005two}
	Wu, Yabo and Dong, Peng and Deng, Xuemei and Zhao, Guoming, The two NUT-like solutions of Ernst equation, \hyperref{https://pubs.aip.org/aip/jmp/article-abstract/46/5/052502/925834/The-two-NUT-like-solutions-of-Ernst-equation?redirectedFrom=fulltext}{}{}{\color{blue}{Journal of mathematical physics. 46 (2005)}}.
	
	
	
	\bibitem{Ernst1968-3}
	J. Ehlers, Konstruktionen und Charakterisierung von Lösungen der Einsteinschen Gravitationsfeld Gleichungen, \hyperref{https://inspirehep.net/literature/45502}{}{}{\color{blue}{dissertation, Hamburg. (1957)}}.
	
	
	
	\bibitem{Ernst1968-4}
	B. Kent Harrison; New Solutions of the Einstein‐Maxwell Equations from Old, \hyperref{https://doi.org/10.1063/1.1664508}{}{}{\color{blue}{J. Math. Phys. 9, 1744-1752 (1968)}}.
	
	\bibitem{Ernst1976}	
	F. J. Ernst; Black holes in a magnetic universe, \hyperref{https://doi.org/10.1063/1.522781}{}{}{\color{blue}{J. Math. Phys. 17, 54-56 (1976)}}.
	
	
	
	
	
	\bibitem{Misner1967}
	C. W. Misner. Taub-NUT Space as a Counterexample to almost anything, \hyperref{https://core.ac.uk/download/pdf/85252979.pdf}{}{}{\color{blue}{Relativity Theory and Astrophysics. 1, 167 (1967)}}.
	
	
	\bibitem{Chng2006}
	B. Chng, R. Mann, and C. Stelea. Accelerating Taub-NUT and Eguchi-Hanson solitons in four dimensions, \hyperref{https://link.aps.org/doi/10.1103/PhysRevD.74.084031}{}{}{\color{blue}{Physical Review D 74. 084031 (2006)}}, \hyperref{https://arxiv.org/abs/gr-qc/0608092v1}{}{}{\color{blue}{arXiv:gr-qc/0608092}}.
	
	
	\bibitem{astorino2013embedding}
	Astorino, Marco, Embedding hairy black holes in a magnetic universe, \hyperref{https://journals.aps.org/prd/abstract/10.1103/PhysRevD.87.084029}{}{}{\color{blue}{Physical Review D—Particles, Fields, Gravitation, and Cosmology. 87, 084029 (2013)}}, \hyperref{https://arxiv.org/abs/1301.6794v2}{}{}{\color{blue}{arXiv:1301.6794 [gr-qc]}}.
	
	
	\bibitem{astorino2020enhanced}
	Astorino, Marco, Enhanced Ehlers transformation and the Majumdar-Papapetrou-NUT spacetime, \hyperref{https://link.springer.com/article/10.1007/JHEP01(2020)123}{}{}{\color{blue}{Journal of High Energy Physics. 2020, 1--38 (2020)}}, \hyperref{https://arxiv.org/abs/1906.08228}{}{}{\color{blue}{arXiv:1906.08228 [gr-qc]}}.
	
	
	
	\bibitem{barrientos2023ehlers}
	Barrientos, Jos{\'e} and Cisterna, Adolfo, Ehlers transformations as a tool for constructing accelerating NUT black holes, \hyperref{https://journals.aps.org/prd/abstract/10.1103/PhysRevD.108.024059}{}{}{\color{blue}{Physical Review D. 108, 024059 (2023)}}, \hyperref{https://arxiv.org/abs/2305.03765v4}{}{}{\color{blue}{arXiv:2305.03765 [gr-qc]}}.
	
	
	\bibitem{barrientos2023pleban}
	Barrientos, Jos{\'e} and Cisterna, Adolfo and Pallikaris, Konstantinos, Pleban'ski-Demia'nskia la Ehlers-Harrison: Exact Rotating and Accelerating Type I Black Holes, \hyperref{https://arxiv.org/abs/2309.13656v2}{}{}{\color{blue}{arXiv:2309.13656 [gr-qc]}}.
	
	
	
	\bibitem{Cisterna2023}
	A. Cisterna, K. Müller, K. Pallikaris, and A. Viganò, Exact rotating wormholes via Ehlers transformations, \hyperref{https://link.aps.org/doi/10.1103/PhysRevD.108.024066}{}{}{\color{blue}{Phys. Rev. D. 108, 024066 (2023)}}, \hyperref{https://arxiv.org/abs/2306.14541}{}{}{\color{blue}{arXiv:2306.14541 [gr-qc]}}. 
	
	
	
	\bibitem{duan1979TSU}
	 Duan, YS and Ge, Mo-Lin; TSU (2) gauge theory and electrodynamics with N magnetic monopoles, \hyperref{https://inspirehep.net/literature/1662393}{}{}{\color{blue}{Sci. Sin. 9, 1072 (1979)}}.
	
	
	
	\bibitem{duan1984structure}
	Duan, YS; The structure of the topological current, \hyperref{https://www.slac.stanford.edu/pubs/slacpubs/3250/slac-pub-3301.pdf}{}{}{\color{blue}{Preprint SLAC-PUB-3301/84. (1984)}}.
	
	
	
	\bibitem{Duan2000}
	Y. Duan, L. Fu, G. Jia; Topological tensor current of $\tilde{p}$-branes in the $\Phi$-mapping theory, \hyperref{https://doi.org/10.1063/1.533347}{}{}{\color{blue}{J. Math. Phys. 41, 4379-4386 (2000)}}, \hyperref{https://arxiv.org/abs/hep-th/9904123}{}{}{\color{blue}{arXiv:hep-th/9904123}}.
	
	
	
	
	\bibitem{Bekenstein1947}
	J. Bekenstein, Exact solution of Einstein conformal scalar equations, \hyperref{https://doi.org/10.1016/0003-4916(74)90124-9}{}{}{\color{blue}{Ann. Phys. (N.Y.) 82, 535 (1974)}}.
	
	
	\bibitem{Ehlers1962}
	J. Ehlers, Transformations of static exterior solutions of Einstein’s gravitational field equations into different solutions by means of conformal mapping, 	\hyperref{https://inspirehep.net/literature/45503}{}{}{\color{blue}{Colloq. Int. 91, 275–284 (1962)}}.
	
	
	\bibitem{Momeni2005}
	D. Momeni, M. Nouri-Zonoz, and R. Ramezani-Arani, Morgan-Morgan-NUT disk space via Ehlers transformation, 	\hyperref{https://link.aps.org/doi/10.1103/PhysRevD.72.064023}{}{}{\color{blue}{Phys. Rev. D. 72, 064023 (2005)}}, \hyperref{https://www.arxiv.org/abs/gr-qc/0508036}{}{}{\color{blue}{arXiv:gr-qc/0508036}}.
	
	
	
	
	
	\bibitem{Alawadhi2020}
	R. Alawadhi, D.S. Berman, B. Spence, et al. S-duality and the double copy, \hyperref{https://link.springer.com/article/10.1007/JHEP03(2020)059}{}{}{\color{blue}{J. High Energ. Phys. 2020, 1-27 (2020)}}, \hyperref{https://arxiv.org/abs/1911.06797}{}{}{\color{blue}{arXiv:1911.06797 [hep-th]}}.
	
	
	
	\bibitem{tipler1976causality}
	Tipler, Frank J, Causality violation in asymptotically flat space-times, \hyperref{https://journals.aps.org/prl/abstract/10.1103/PhysRevLett.37.879}{}{}{\color{blue}{Physical Review Letters. 37, 879 (1976)}}.
	
	
	
	\bibitem{visser1995lorentzian}
	Visser, Matt, Lorentzian wormholes. from Einstein to Hawking, \hyperref{https://ui.adsabs.harvard.edu/abs/1995lwet.book.....V/abstract}{}{}{\color{blue}{Woodbury. 289-296 (1995)}}.
	
	
	
    \bibitem{rm1984general}
	RM, Wald, General Relativity, \hyperref{https://press.uchicago.edu/ucp/books/book/chicago/G/bo5952261.html}{}{}{\color{blue}{University of Chicago Press (1984)}}.
	
	
	
	\bibitem{lobo2010closed}
	Lobo, Francisco SN, Closed timelike curves and causality violation, \hyperref{https://inspirehep.net/literature/864696}{}{}{\color{blue}{Classical and Quantum Gravity. 19 (2008)}}, \hyperref{https://arxiv.org/abs/1008.1127v1}{}{}{\color{blue}{arXiv:1008.1127 [gr-qc]}}.
	


	
	
	\bibitem{Ye2023}
	X. Ye, and S. Wei, Distinct topological configurations of equatorial timelike circular orbit for spherically symmetric (hairy) black holes, \hyperref{https://iopscience.iop.org/article/10.1088/1475-7516/2023/07/049}{}{}{\color{blue}{Journal of Cosmology and Astroparticle Physics 2023, 049 (2023)}}, \hyperref{https://arxiv.org/abs/2301.04786}{}{}{\color{blue}{arXiv:2301.04786 [gr-qc]}}.
	
	
	
	
	
	\bibitem{Cunha2016}
	P. V. P. Cunha, J. Grover, C. Herdeiro, E. Radu, H. Runarsson and A. Wittig, Chaotic lensing around boson stars and Kerr black holes with scalar hair, \hyperref{https://doi.org/10.1103/PhysRevD.94.104023}{}{}{\color{blue}{Phys. Rev. D. 94, 104023 (2016)}}, \hyperref{https://arxiv.org/abs/1609.01340}{}{}{\color{blue}{arXiv:1609.01340 [gr-qc]}}.
	
	
	\bibitem{Cunha2020}
	P. V. P. Cunha, and C. A. R. Herdeiro, Stationary black holes and light rings, \hyperref{https://link.aps.org/doi/10.1103/PhysRevLett.124.181101}{}{}{\color{blue}{Phys. Rev. Lett. 124, 181101 (2020)}}, \hyperref{https://doi.org/10.48550/arXiv.2003.06445}{}{}{\color{blue}{arXiv:2003.06445[gr-qc]}}.
	
	\bibitem{Wei2020}
	S. W. Wei, Topological Charge and Black Hole Photon Spheres, \hyperref{https://link.aps.org/doi/10.1103/PhysRevD.102.064039}{}{}{\color{blue}{Phys. Rev. D. 102, 064039 (2020)}}, \hyperref{https://doi.org/10.48550/arXiv.2006.02112}{}{}{\color{blue}{arXiv:2006.02112[gr-qc]}}.
	
	
	
	
	\bibitem{Guo2021}
	M. Guo and S. Gao, Universal Properties of Light Rings for Stationary Axisymmetric Spacetime, \hyperref{https://link.aps.org/doi/10.1103/PhysRevD.103.104031}{}{}{\color{blue}{Phys. Rev. D. 103, 104031 (2021)}}, \hyperref{https://doi.org/10.48550/arXiv.2011.02211}{}{}{\color{blue}{arXiv:2011.02211[gr-qc]}}.
	
	
	
	\bibitem{Duan1984}
	Y. S. Duan, The structure of the topological current,
	\hyperref{https://www.slac.stanford.edu/pubs/slacpubs/3250/slac-pub-3301.pdf}{}{}{\color{blue}{Phys. Rev. D. (1984)}}.
	
	
	\bibitem{Cunha2017}
	P. V. P. Cunha, E. Berti, and C. A. R. Herdeiro, Light ring stability in ultra-compact objects, \hyperref{https://link.aps.org/doi/10.1103/PhysRevLett.119.251102}{}{}{\color{blue}{Phys. Rev. Lett. 119, 251102 (2017)}}, \hyperref{https://arxiv.org/abs/1708.04211}{}{}{\color{blue}{arXiv:1708.04211 [gr-qc]}}.
	
	
	
	
	
	
	
	\bibitem{Wei2022}
	S. W. Wei and Y. X. Liu, Topology of equatorial timelike circular orbits around stationary black holes, \hyperref{https://link.aps.org/doi/10.1103/PhysRevD.107.064006}{}{}{\color{blue}{Phys. Rev. D. 107, 064006 (2023)}}, \hyperref{https://doi.org/10.48550/arXiv.2207.08397}{}{}{\color{blue}{arXiv:2207.08397[gr-qc]}}.
	
	
	
	\bibitem{ferrari1984oscillations}
	V. Ferrari and B. Mashhoon, Oscillations of a black hole, \hyperref{https://link.aps.org/doi/10.1103/PhysRevLett.52.1361}{}{}{\color{blue}{}Phys. Rev. Lett. 52, 1361 (1984)}.
	
	\bibitem{ferrari1984new}
	V. Ferrari and B. Mashhoon, New approach to the quasinormal modes of a black hole, \hyperref{https://link.aps.org/doi/10.1103/PhysRevD.30.295}{}{}{\color{blue}{Phys. Rev. D. 30, 295 (1984)}}.
	
	\bibitem{mashhoon1985stability}
	B. Mashhoon, Stability of charged rotating black holes in the eikonal approximation, \hyperref{https://link.aps.org/doi/10.1103/PhysRevD.31.290}{}{}{\color{blue}{Phys. Rev. D. 31, 290 (1985)}}.
	
	
	\bibitem{schneider1992gravitational}
	Schneider, Peter and Ehlers, J{\"u}rgen and Falco, Emilio E and Schneider, Peter and Ehlers, J{\"u}rgen and Falco, Emilio E, Gravitational lenses as astrophysical tools, \hyperref{https://link.springer.com/book/10.1007/3-540-45857-3}{}{}{\color{blue}{Gravitational Lenses. 467--515 (1992)}}.
	
	
	
	\bibitem{wambsganss1998gravitational}
	Wambsganss, Joachim, Gravitational lensing in astronomy, \hyperref{https://link.springer.com/article/10.12942/lrr-1998-12}{}{}{\color{blue}{Living Reviews in Relativity. 1, 1--74 (1998)}}.
	
	
	
	\bibitem{narayan1999gravitational}
	Narayan, Ramesh and Bartelmann, Matthias, Gravitational lensing, \hyperref{https://ui.adsabs.harvard.edu/abs/1999fsu..conf..360N/abstract}{}{}{\color{blue}{Formation of Structure in the Universe. 360 (1999)}}.
    
    
    \bibitem{schneider2006gravitational}
    Schneider, Peter and Kochanek, Christopher and Wambsganss, Joachim, Gravitational lensing: strong, weak and micro: Saas-Fee advanced course 33, \hyperref{https://link.springer.com/book/10.1007/978-3-540-30310-7}{}{}{\color{blue}{Springer Science \& Business Media. 33 (2006)}}.
    
    
    \bibitem{gibbons2008applications}
    Gibbons, GW and Werner, MC, Applications of the Gauss--Bonnet theorem to gravitational lensing, \hyperref{https://iopscience.iop.org/article/10.1088/0264-9381/25/23/235009/meta}{}{}{\color{blue}{Classical and Quantum Gravity. 25, 235009 (2008)}}, \hyperref{https://arxiv.org/abs/0807.0854v1}{}{}{\color{blue}{arXiv:0807.0854 [gr-qc]}}.
    
    
   
    

	
	
	
	
	
	
	
	
	
	
	
	
	
	
	
	
	
	
	
	
\end{thebibliography}
\end{document}